\documentclass{aa}  

\usepackage{graphicx}
\usepackage{natbib} 
\bibpunct{(}{)}{;}{a}{}{,} 

\usepackage[varg]{txfonts}

\newcommand{\kms}{km\,s$^{-1}$}
\newcommand{\um}{$\mu$m}

\newcommand{\cmc}{cm$^{-3}$}

\newcommand{\msol}{M$_{\odot}$}
\newcommand{\msolyr}{M$_{\odot}$\,yr$^{-1}$}

\begin{document}
   \title{The jet and the disk of the HH 212 low-mass protostar\\
imaged by ALMA: SO and SO$_2$ emission}

   \author{L. Podio \inst{1} \and
     C. Codella \inst{1} \and
     F. Gueth  \inst{2} \and
     S. Cabrit \inst{3,4} \and
     R. Bachiller \inst{5} \and
     A. Gusdorf \inst{3} \and
     C.-F. Lee \inst{6} \and
     B. Lefloch \inst{4} \and
     S. Leurini \inst{7} \and
     B. Nisini \inst{8} \and
     M. Tafalla \inst{5}
          }

   \institute{
     INAF - Osservatorio Astrofisico di Arcetri, Largo E. Fermi 5, 50125 Firenze, Italy \\ 
     \email{lpodio@arcetri.astro.it}
     \and
     IRAM, 300 rue de la Piscine, 38406 Saint Martin d'H\`eres, France
     \and
     LERMA, Observatoire de Paris, UPMC Univ. Paris 06, PSL Research University, Sorbonne Universit\'es, CNRS, F-75014, Paris, France
     \and
     UJF-Grenoble1/CNRS-INSU, Institut de Plan\'etologie et d'Astrophysique de Grenoble (IPAG) UMR 5274, Grenoble, 38041, France
     \and
     IGN, Observatorio Astron\'omico Nacional, Alfonso XIII 3, 28014, Madrid, Spain
     \and
     Academia Sinica Institute of Astronomy and Astrophysics, P.O. Box 23-141, Taipei 106, Taiwan
     \and
     Max-Planck-Institut f\"ur Radioastronomie, Auf dem H\"ugel 69, 53121 Bonn, Germany
     \and
     INAF, Osservatorio Astronomico di Roma, via di Frascati 33, 00040, Monte Porzio Catone, Italy
             }

   \date{Received ; accepted }

 
  \abstract
{To investigate the disk formation and jet launching mechanism in protostars is crucial to comprehend the earliest stages of star and planet formation. 
}
{We aim to constrain the physical and dynamical properties of the molecular jet and the disk of the HH~212 protostellar system at unprecedented angular scales exploiting the capabilities of the Atacama Large Millimeter Array (ALMA).}
{ALMA observations of HH~212 in emission lines from sulfur-bearing molecules, SO~$9_8-8_7$, SO~$10_{11}-10_{10}$, SO$_2$~$8_{2,6}-7_{1,7}$, are compared with simultaneous CO~$3-2$, SiO~$8-7$ data. 
The molecules column density and abundance are estimated using simple radiative transfer models.}
{SO~$9_8-8_7$ and SO$_2$~$8_{2,6}-7_{1,7}$ show broad velocity profiles. At systemic velocity they probe the circumstellar gas and the cavity walls. Going from low to high blue-/red-shifted velocities the emission traces the wide-angle outflow and the fast ($\sim100-200$~\kms) and collimated ($\sim90$~AU) molecular jet revealing the inner knots with timescales $\le50$ years. The jet transports a mass loss rate $\ge0.2-2\times10^{-6}$~\msolyr, implying high ejection efficiency ($\ge0.03-0.3$). The SO and SO$_2$ abundances in the jet are $\sim10^{-7}-10^{-6}$.
SO~$10_{11}-10_{10}$ emission is compact and shows small-scale velocity gradients indicating that it originates partly from the rotating disk previously seen in HCO$^{+}$ and C$^{17}$O, and partly from the base of the jet. 
The disk mass is $\ge0.002-0.013$~\msol\, and the SO abundance in the disk is $\sim10^{-8}-10^{-7}$.}
{SO and SO$_2$ are effective tracers of the molecular jet in the inner few hundreds AU from the protostar.
Their abundances indicate that $1\%-40\%$ of sulfur is in SO and SO$_2$ due to shocks in the jet/outflow and/or to ambipolar diffusion at the wind base.
The SO abundance in the disk is $3-4$ orders of magnitude larger than in evolved protoplanetary disks. This may be due to an SO enhancement in the accretion shock at the envelope-disk interface or in spiral shocks if the disk is partly gravitationally unstable.}

\keywords{Stars: formation -- ISM: jets \& outflows -- ISM: molecules -- ISM: individual: HH 212 
               }

               \maketitle
%

\section{Introduction}

\begin{figure*}
  \begin{centering}
  \includegraphics[width=12.cm]{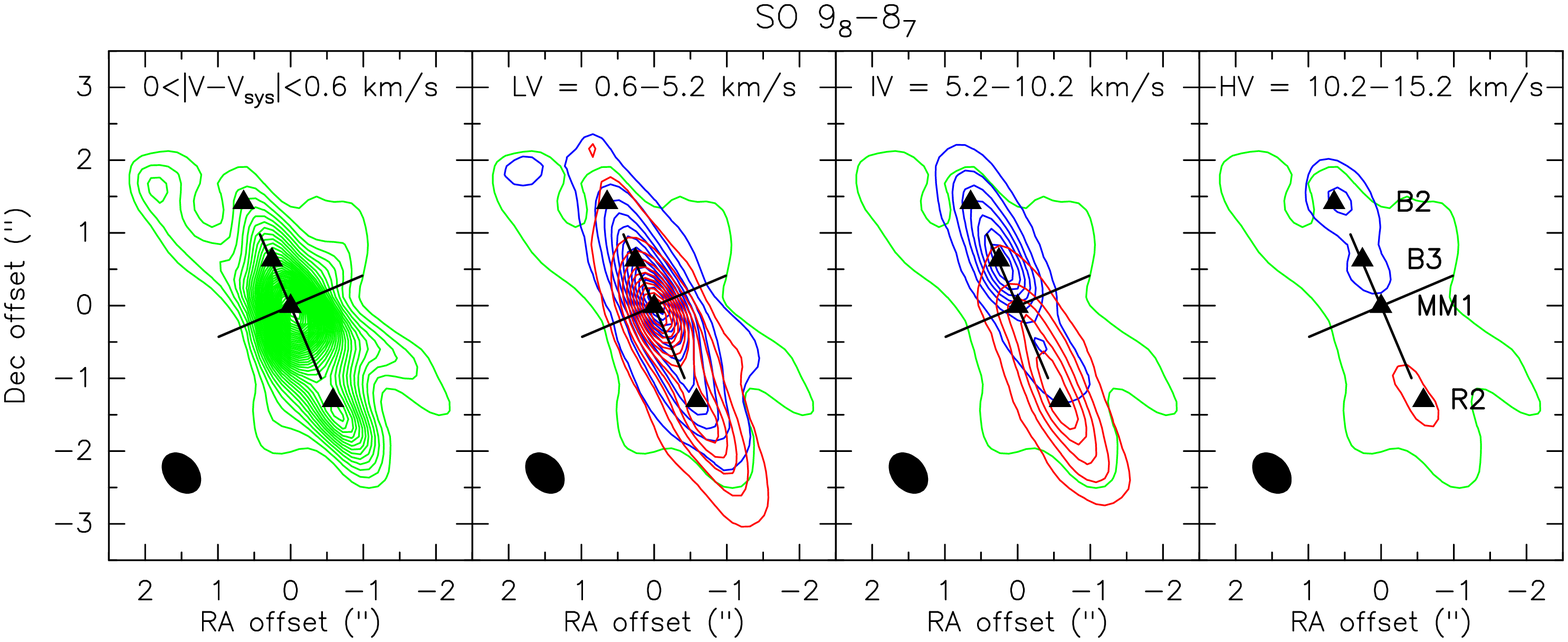}
  \includegraphics[width=12.cm]{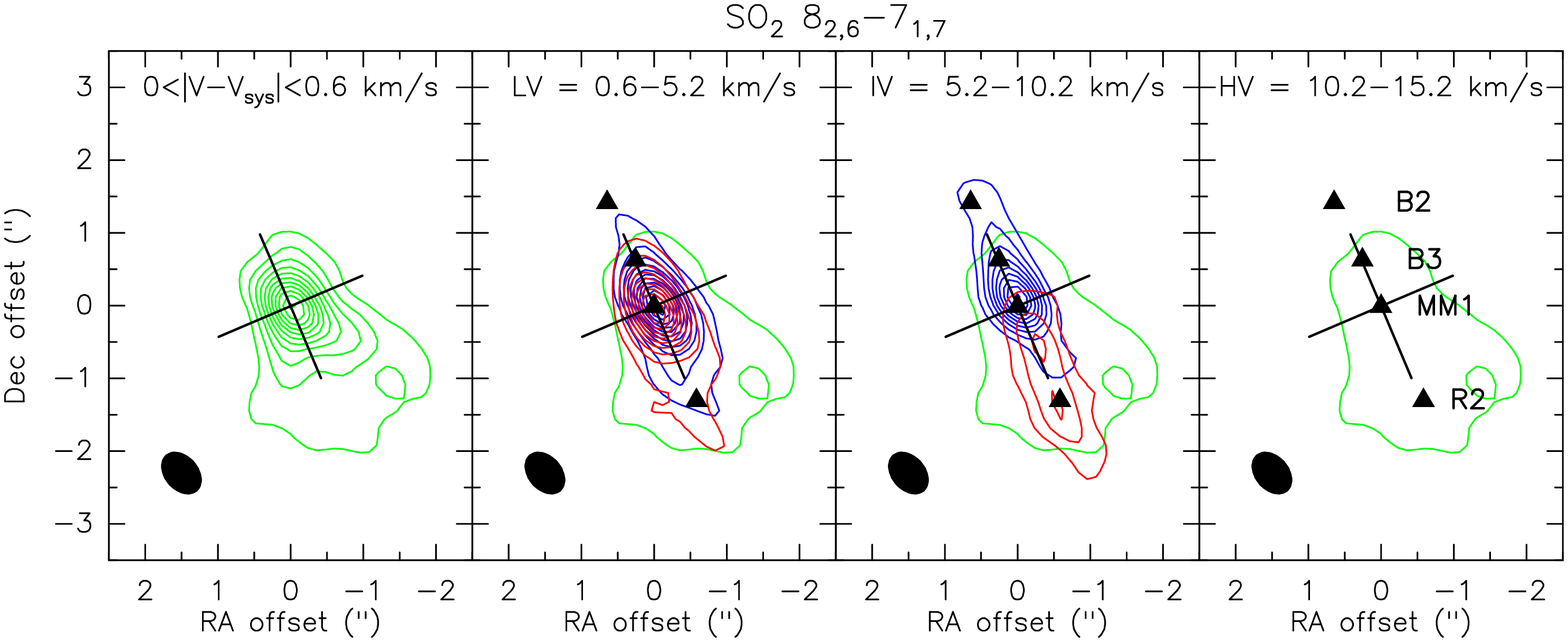}
  \caption{Channel maps of SO $9_8-8_7$ ({\it top}) and SO$_2$ $8_{2,6}-7_{1,7}$ ({\it bottom}) at systemic, low (LV), intermediate (IV), and high (HV) velocities ({\it panels from left to right}).  
The green, blue, and red contours trace the emission at systemic velocity and over the labeled blue- and red-shifted velocity intervals ($| V - V_{\rm sys} |$). The lowest contour of the emission at systemic velocity is shown in all panels. The first contour is at 5$\sigma$ for SO and 3$\sigma$ for SO$_2$ with steps of 5$\sigma$ ($V_{\rm sys}$) and 20$\sigma$ (LV, IV, HV) for SO and 3$\sigma$ for SO$_2$ (1$\sigma \sim$ 3-4 mJy/beam/0.43 \kms). The tilted black cross is centred at the source position and indicates the jet and the disk PA (PA$_{\rm jet} = 22\degr$, PA$_{\rm disk} = 112\degr$). The ellipse in the bottom-left corner shows the beam HPBW of the line emission maps ($0\farcs65 \times 0\farcs47$, PA = 49$\degr$). The triangles indicate the emission peaks along the blue and red lobes.}
  \label{fig:SO_vel}
  \end{centering}
\end{figure*}

The first steps of the formation of a low-mass star are regulated by the simultaneous effects of the mass accretion onto the star and the ejection of matter from the stellar-disk system.
As a result, protostars (the so-called Class 0 objects) are characterised by the occurrence of
fast bipolar jets flowing perpendicular to the plane of accretion disks. 
In practice, although the precise launch region (star, inner disk edge at $\sim$ 0.1 AU, outer disk at $\sim$ 0.1--10 AU) remains unknown \citep[e.g., ][]{ferreira06}, 
jets are thought to remove excess angular momentum from the star-disk system, thus allowing disk accretion onto the central object.
Unfortunately, the observations of the jet-disk pristine systems in deeply embedded protostars are very difficult to perform
given the small scales involved as well as the occurrence of numerous other kinematical components involved
in the star formation recipe (cavities of swept-up material, infalling envelope, static ambient cloud).                                      

The HH 212 region in Orion (at 450 pc) can be considered as an ideal laboratory to investigate the interplay
of infall, outflow and rotation in the earliest evolutionary phases of the star forming process. 
HH 212 is low-mass Class 0 source driving a symmetric                                   
and bipolar jet extensively observed in typical molecular tracers such as H$_2$, SiO, and CO \citep[e.g., ][]{zinnecker98}.
High-spatial resolution observations (down to $\simeq 0\farcs3 - 0\farcs4$) performed with the SubMillimeter Array (SMA) \citep{lee06,lee07a,lee08}, 
the IRAM Plateau de Bure (PdB) interferometer \citep{codella07,cabrit07b,cabrit12}, and the Atacama Large Millimeter Array (ALMA; see \citealt{lee14,codella14b})   
reveal the inner $\pm 1\arcsec - 2\arcsec = 450-900$ AU collimated  jet (width $\simeq$ 100 AU) close to the protostar.  
HH 212 is also associated with a flattened rotating envelope in the equator perpendicular to the jet axis observed
firstly with the NRAO Very Large Array (VLA) in NH$_3$ emission by \citet{wiseman01} on 6000 AU scales.
More recent SMA and ALMA observations in the CO isotopologues and HCO$^{+}$ on  $\sim 2000$, $800$ AU scales indicates that the flattened envelope is not only rotating but also infalling onto the central source, and can therefore be identified as a pseudo-disk according to magnetized core collapse models \citep{lee06,lee14}.
In addition, a compact ($\le 120$ AU), optically thick dust peak is observed by \citet{codella07,lee06,lee14} and attributed to an edge-on disk rather than the inner envelope.
This seems to be confirmed by HCO$^{+}$ and C$^{17}$O emission showing signatures of a compact disk of radius $\sim$ 90 AU keplerian rotating around a source of  
$\simeq 0.2 - 0.3$ \msol\,\citep{lee14,codella14b}.
Keplerian rotating disks had previously been observed towards only other three Class 0 objects:
 IRAS 4A2 \citep{choi10}, L1527 \citep{tobin12,sakai14}, and VLA1623A \citep{murillo13} but HH 212 can be
considered as the only object clearly
revealing {\it both} a disk and a fast collimated jet, calling for further observations aimed to characterise
its inner regions.

Recent observations show that SO can be used to image high-velocity
protostellar jets, similarly to a standard tracer such as SiO \citep{lee10,tafalla10,codella14a}.
However, only two protostellar jets have been so far clearly mapped in SO lines: HH 211 and NGC1333-IRAS2A,
and estimates of the SO abundance have been derived only for HH 211.
On the other hand, sulfur bearing species have been long searched in protoplanetary disks associated with evolved young
stellar objects (i.e. Class I-II) but only CS has been routinely 
observed while SO is hard to detect: 
\citet{dutrey11} reported no detection of SO (and H$_2$S) from three prototypical protoplanetary disks, 
\citet{fuente10} reported a detection in the disk of the T Tauri star AB Aur, and \citet{guilloteau13} showed that SO is exceptionally observed in disks with only one definite detection on a sample of 42 T Tauri and Herbig Ae stars towards 04302+2247.       
The statistics regarding disks around Class 0 protostars is even poorer.
The only case is represented by L1527, where \citet{sakai14} have shown that SO as observed by ALMA originates from the outer disk near the centrifugal barrier, and its emission is argued to be enhanced by an accretion shock.
Observations of HH 212 in the SO $9_8-8_7$ line
obtained with the SMA by \citet{lee07a} 
suggest that SO may have two components, a low-velocity one originating in the inner rotating envelope/pesudo-disk and an high-velocity one from the jet. 
However, a detailed analysis of these two components was prevented by the lack of angular resolution and sensitivity.

In this paper we exploit the unprecedented combination of high-spatial resolution and high-sensitivity of ALMA to image and characterise both the molecular jet and 
the disk around the HH 212 protostar through SO and SO$_2$ lines.
In order to evaluate the reliability of SO and SO$_2$ lines as tracers of shock chemistry and/or high-density gas in the disk, the SO and SO$_2$ spatio-kinematical properties are compared with a well-know shock chemistry tracer (SiO), with a universal outflow tracer not sensitive to density or chemistry (CO), and with a disk tracer (C$^{17}$O).
The ALMA observations of SO and SO$_2$ emission presented in Sect. \ref{sect:observations} are at higher angular resolution than previous SMA ones by \citet{lee07a} (HPBW $= 0\farcs65 \times 0\farcs47$, i.e. around a factor 4 better than at SMA, HPBW $= 1\farcs16 \times 0\farcs84$)
and $\sim 100$ times more sensitive ($\sigma \sim 3-4$ mJy/beam/0.43 \kms\ with ALMA and $\sigma \sim 450$ mJy/beam/\kms\ with SMA). 
This allows us to disentangle for the first time the origin of the different velocity components in SO and SO$_2$ and to probe the different structures in the compact circumstellar region, i.e. the cavity walls, the outflow, the molecular jet, the envelope, and the disk (see Sect. \ref{sect:results}). 
The simultaneous observation of the SiO $8-7$ and CO/C$^{17}$O $3-2$ lines presented in \citet{codella14b} allows determining the molecules abundances and the physical and dynamical properties of the jet and the disk (see Sect. \ref{sect:discussion}).
Finally, our conclusions are summarized in Sect. \ref{sect:conclusions}.

\section{Observations and data reduction}
\label{sect:observations}

\begin{table}
  \caption[]{\label{tab:lines} Properties of the observed transitions.}
  \begin{tabular}[h]{ccccc}
    \hline
    \hline
    Line & $\nu_{0}$ & $E_{\rm up}$ & $S_{\rm ij} \, \mu^2$ & log$_{10} A_{\rm ij}$ \\
    & (MHz)      & (K)       & (D$^2$)  & (s$^{-1}$)     \\
    \hline 
    \hline
    SO$_2$ $8_{2,6}-7_{1,7}$ & 334673.352 & 43 & 4.9 & $-3.9$  \\
    SO $10_{11}-10_{10}$ & 336553.346 & 143 & 0.3 & $-5.2$ \\ 
    C$^{17}$O $3-2$ & 337061.130  & 32 & 0.04 & $-5.6$ \\
    CO $3-2$ & 345795.990 & 33 & 0.04 & $-5.6$ \\
    SO $9_8-8_7$ & 346528.481 & 79 & 21.5 & $-3.3$ \\
    SiO $8-7$ & 347330.631 & 75 & 77.0 & $-2.7$ \\
    \hline     
  \end{tabular}\\
\small
$^{a}$from the JPL molecular database \citep{pickett98}
\end{table}

\begin{figure*}
  \begin{centering}
  \includegraphics[width=\textwidth]{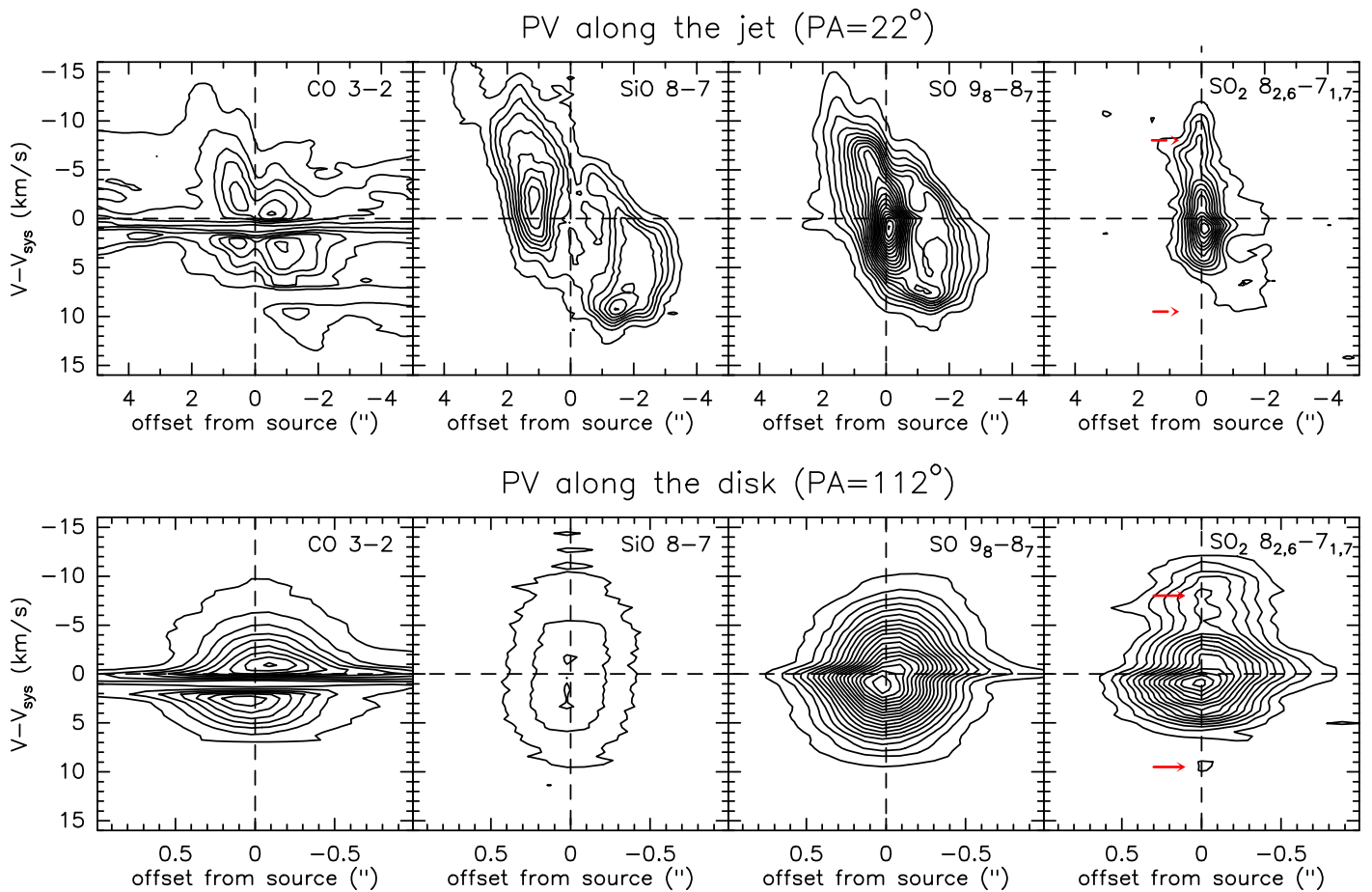}
  \caption{{\it From left to right:} Position-velocity diagram of CO $3-2$, SiO $8-7$, SO $9_8-8_7$, and SO$_2$ $8_{2,6}-7_{1,7}$ lines obtained along the jet (PA$_{\rm jet} = 22\degr$, {\it top}) and the disk (PA$_{\rm disk} = 112\degr$, {\it bottom}). Horizonthal and vertical dashed lines mark V$_{\rm sys} = +1.3$ \kms\, and the continuum peak MM1.
The red arrows on the SO$_2$ $8_{2,6}-7_{1,7}$ PV indicates emission from complex organic molecules which is blended with the SO$_2$ emission at the source position.
For CO $3-2$, SiO $8-7$, SO $9_8-8_7$, SO$_2$ $8_{2,6}-7_{1,7}$ the first contour is at 5$\sigma$ ($\sigma = 10$ mJy for SiO, 4 mJy for SO, and 3 mJy for SO$_2$) with steps of 10$\sigma$ (CO, SiO, and SO), and 5$\sigma$ (SO$_2$).}
  \label{fig:pv}
  \end{centering}
\end{figure*}

HH 212 was observed in Band 7 in the extended configuration of the ALMA Early Science Cycle 0 operations on December 1 2012 using 24 antennas of 12-m. The shortest baseline was about 20 m and the longest one 360 m, hence the maximum unfiltred scale is of 3$\arcsec$ at 850\um.
The properties of the observed SO 9$_8$-8$_7$, SO $10_{11}-10_{10}$, and SO$_2$ $8_{2,6}-7_{1,7}$ transitions are summarized in Table \ref{tab:lines} (frequency $\nu_0$ in MHz, upper level energy $E_{\rm up}$ in K, $S_{\rm ij} \, \mu^2$ in D$^2$, coefficient for radiative decay $A_{\rm ij}$ in s$^{-1}$). We also report the properties of the SiO $8-7$, CO $3-2$, and C$^{17}$O $3-2$ transitions observed during the same run and presented by \citet{codella14b}, which are also analysed to compare with the SO and SO$_2$ emission. The obtained datacubes have a spectral resolution of 488 kHz ($0.42-0.43$ \kms), a typical beam FWHM of $0\farcs65 \times 0\farcs47$ at PA$\sim 49\degr$, and an rms noise of $\sim 3-4$ mJy/beam in the 0.43 \kms\ channel. 
The calibration was carried out following standard procedures and using quasars J0538–440, J0607–085, as well as Callisto and Ganymede. Spectral line imaging and data analysis were performed using the CASA\footnote{http://casa.nrao.edu} and the GILDAS\footnote{http://www.iram.fr/IRAMFR/GILDAS} packages.
Offsets are given with respect to the MM1 protostar position as determined from the dust continuum peak by \citet{codella14b}, i.e. $\alpha$(J2000) = 05$^h$ 43$^m$ 51$^s$.41, $\delta$(J2000) = $-01\degr$ 02$\arcmin$ 53$\farcs$17. These values are in excellent agreement with those determined through previous SMA and PdBI observations.
Velocities are given with respect to the systemic velocity, $V_{\rm sys}$, which is estimated following the same method as in \citet{codella14b}.
They show that the emission in the C$^{17}$O $3-2$ and C$^{34}$S $7-6$ lines is most extended and most symmetric in the 0.43 \kms\, wide velocity bin centered at  $+ 1.13 \pm 0.22$ \kms\, and $+ 1.42 \pm 0.22$ \kms, respectively, and assume that $V_{\rm sys}$ is the average of these values. For SO 9$_8$-8$_7$ and SO$_2$ $8_{2,6}-7_{1,7}$ the largest extension is observed at $+ 1.27 \pm 0.21$ \kms\, and $+ 1.53 \pm 0.22$ \kms. 
The average of the central velocity of  C$^{17}$O, C$^{34}$S, SO, and SO$_2$ gives $V_{\rm sys} = + 1.3 \pm 0.2$ \kms, in agreement with the value adopted by \citet{codella14b}.
Note that despite the coarse velocity binning, all the lines indicate a systemic velocity significantly smaller than what determined by \citet{wiseman01} ($+ 1.6 \pm 0.1$ \kms) and the value adopted by \citet{lee06,lee07a,lee14} ($+ 1.7 \pm 0.1$ \kms). 
As argued by \citet{codella14b} this discrepancy can be due to the fact that the C$^{17}$O, C$^{34}$S, SO, and SO$_2$ emission seen by ALMA probes the gas motion on much smaller spatial scales ($\le 1350$ AU) than the ammonia observed by \citet{wiseman01} ($\sim 6000$ AU). 
Moreover, SO and SO$_2$ line peaks towards the MM1 protostar are redshifted by $\sim +1$ \kms\, with respect to the velocity where they show the maximum spatial extent (see Sect. \ref{sect:envelope}). This may due to optical depth effets, and suggests that adopting the velocity where the ambient gas show the maximum extent may be more accurate than using the line peak velocity to define $V_{\rm sys}$.
Note, however, that the results presented in the paper are not dependent on the $0.3$ \kms\, velocity difference between the different $V_{\rm sys}$ estimates.


\section{Results}
\label{sect:results}

In order to constrain the origin of the detected SO $9_8-8_7$ and  SO$_2$ $8_{2,6}-7_{1,7}$ lines we define four velocity intervals over which the emission show different morphologies and kinematic properties\footnote{the defined velocity intervals are the same as in \citet{cabrit07b}}:
\begin{itemize}
\item[-] systemic velocity: $0 < |V - V_{\rm sys}| < 0.6$ \kms;   
\item[-] low-velocity (LV hereafter):  $0.6 < |V - V_{\rm sys}| < 5.2$ \kms;
\item[-] intermediate-velocity (IV hereafter): $5.2 < |V - V_{\rm sys}| < 10.2$ \kms;
\item[-] high-velocity (HV hereafter): $10.2 < |V - V_{\rm sys}| < 15.2$ \kms.
\end{itemize}

Figure \ref{fig:SO_vel} shows channel maps of the the SO 9$_8$-8$_7$ and SO$_2$ $8_{2,6}-7_{1,7}$ emission in the four velocity intervals defined above. 
In the following sections we discuss the origin and the properties of the different SO and SO$_2$ velocity components.

\subsection{The cavity walls}
\label{sect:cavity}

\begin{figure}
  \begin{centering}
  \includegraphics[width=\columnwidth]{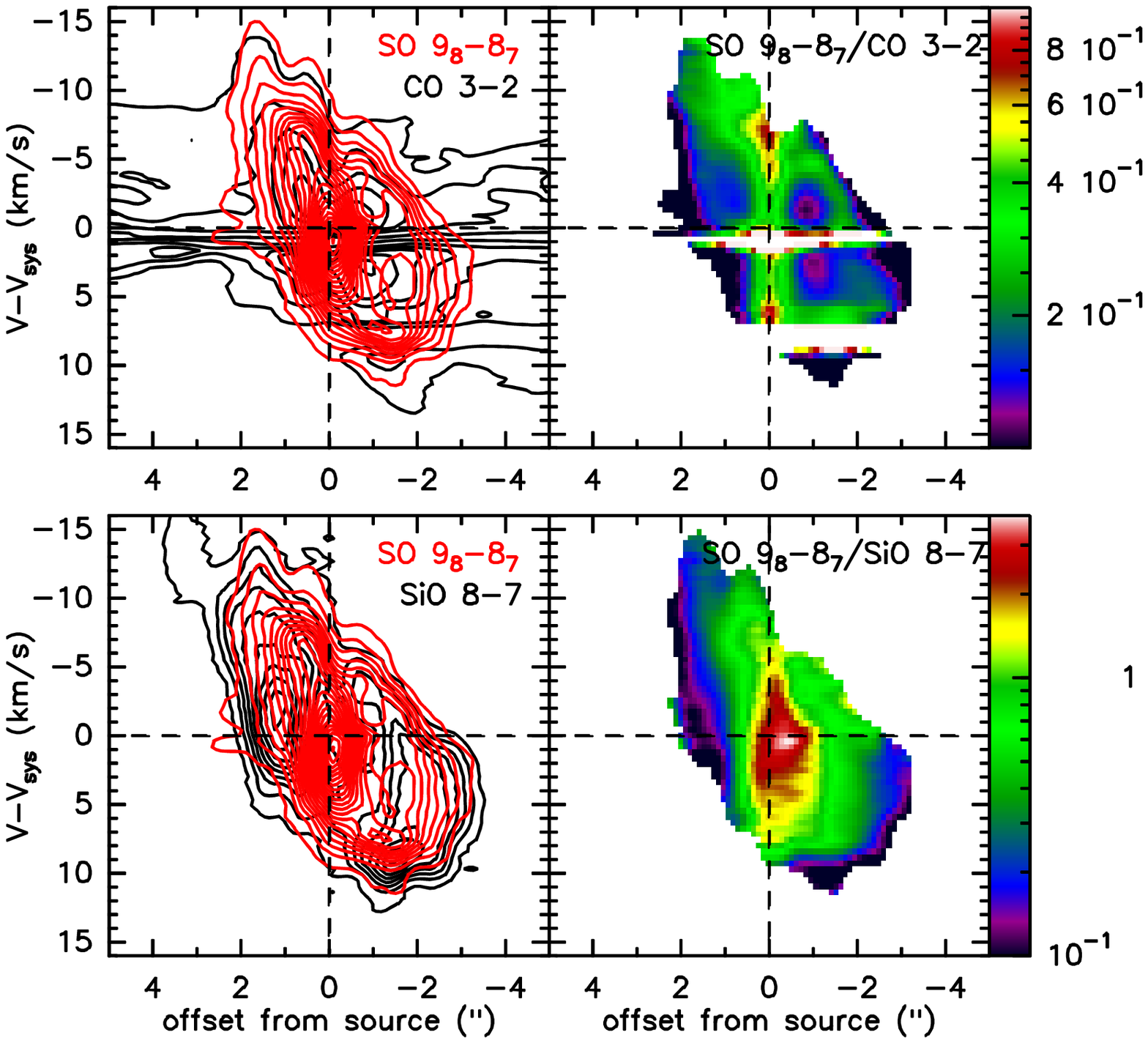}
  \caption{{\it Left panels:} The PV of SO $9_8-8_7$ emission along the jet PA (in red) is overplotted on SiO $8-7$ and CO $3-2$ PVs (in black, top and bottom panels, respectively). {\it Right panels:} Ratio of the SO $9_8-8_7$ PV to the SiO $8-7$ and CO $3-2$ PV (top and bottom panels, respectively). The colour scale is logarithmic.}
  \label{fig:so_vs_co_sio}
  \end{centering}
\end{figure}

At systemic velocity SO 9$_8$-8$_7$ and SO$_2$ $8_{2,6}-7_{1,7}$ peak at the source position (see left panel of Figure \ref{fig:SO_vel}). This may be due to a density enhancement in the circumstellar region, as suggested by the modeling of HCO$^{+}$ $4-3$ emission \citep{lee14}, and/or to an enhancement of the SO and SO$_2$ abundances due to dust grain mantles sublimation and release of sulphur-bearing species in gas-phase. 
The SO 9$_8$-8$_7$ emission also extends in a wide-angle biconical structure around the direction of the H$_2$/SiO jet which extends up to $2\arcsec-3\arcsec$ distance from source and up to $\pm 0.6$ \kms\, with respect to systemic velocity. Also the SO$_2$ emission shows a similar morphology even though only southern to the source. Given the high critical density of the observed SO and SO$_2$ transitions ($n_{\rm cr} \sim 2 \times 10^{6}$ \cmc\, for gas temperatures of $50-100$ K) and the similarity with the C$^{34}$S emission detected by \citet{codella14b}, this emission is believed to originate in the compressed, swept-up gas in the outflow cavity walls.\\

\subsection{The outflow}
\label{sect:outflow}

At low velocities the SO 9$_8$-8$_7$ and SO$_2$ $8_{2,6}-7_{1,7}$ emission is bipolar and collimated along the jet direction. Blue- and red-shifted LV emission largely overlap in the two lobes similarly to SiO $5-4$ and $8-7$ \citep{codella07,codella14b,lee07a}. The emission has a transverse FWHM of $\sim 0\farcs56$, which implies an intrinsic width of $\sim 0\farcs3 = 135$ AU after correction for the ALMA HPBW ($\sim 0\farcs47$ in the transverse direction), i.e. larger than the SiO jet width measured by \citet{cabrit07b} ($\sim 90$ AU for all velocity components). 
This indicates that the LV component, with large blue/red overlap, is probing a slower wider-angle outflow surrounding the narrow HV jet (see the following section).


\subsection{The fast and collimated molecular jet}
\label{sect:jet}

\begin{figure}
  \begin{centering}
  \includegraphics[width=8.cm]{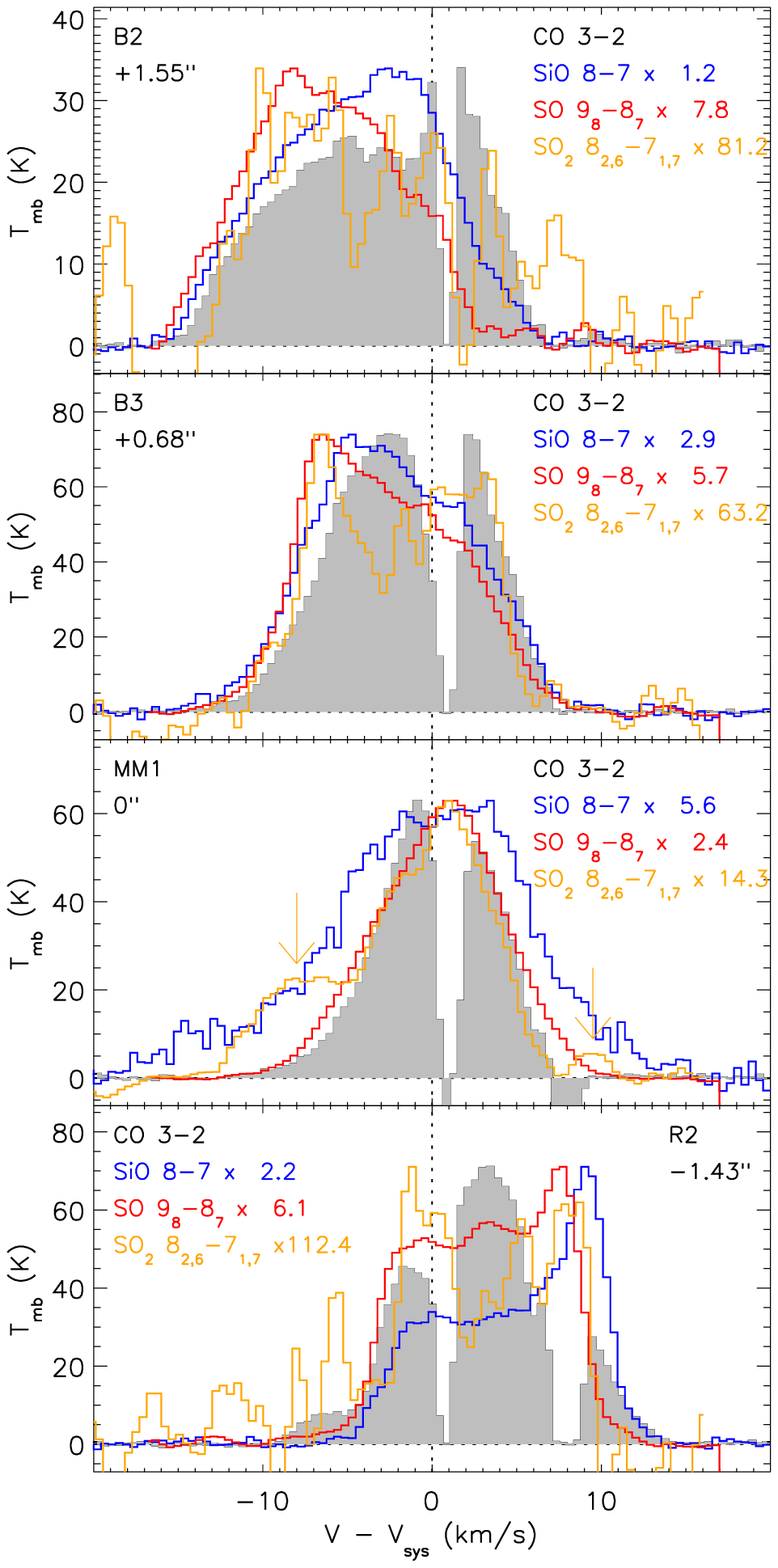}
  \caption{Intensity profiles of SiO $8-7$ (blue),  SO $9_8-8_7$ (red), and SO$_2$ $8_{2,6}-7_{1,7}$ (orange) over CO 3--2 (grey) at the knots (B2, B3, R2) and source (MM1) positions. The name and distance of the knots and the normalization factor to the CO 3--2 peak intensity are indicated in the top of the panels.
The orange vertical arrows at the MM1 position indicate emission lines from complex organic molecules which are blended to the SO$_2$ line.
}
  \label{fig:spec_jet}
  \end{centering}
\end{figure}

\begin{table*}
  \caption[]{\label{tab:fluxes} Properties of the observed emission lines on-source (MM1) and at the position of the knots along the jet.
%
}
  \begin{tabular}[h]{cccccccccc}
    \hline
    \hline
    & rms & $V_{\rm max, b}$$^{a,*}$ & $V_{\rm max, r}$$^{a,*}$ & $FWZI$$^{b,*}$ & $V_{\rm peak}$$^{c,*}$ & $T_{\rm mb, peak}$$^{d}$ & $\int T_{\rm mb} dV$$^{e}$ & HV$_{\rm jet/disk}$$^{f}$ & $\int T_{\rm mb} dV$ (HV$_{\rm jet/disk}$)$^{g}$ \\
    & (K)  & (\kms)          & (\kms)          & (\kms)       & (\kms)           & (K)                 & (K \kms) & (\kms) & (K \kms) \\ 
    \hline 
    \multicolumn{9}{c}{CO $3-2$} \\
    \hline
    B2 &  0.3 & $-16.8$ &   7.3 &  24.1 &   - & 33.9 $\pm$  0.3 &  370.0 $\pm$    1.1 &  $[-8, -15]$ & 82.6 $\pm$   0.5 \\ 
    B3 &  0.3 & $-16.8$ &   7.7 &  24.5 &  - & 73.9 $\pm$  0.3 &  732.5 $\pm$    0.9 &  $[-8, -15]$ & 46.7 $\pm$   0.4 \\ 
    MM1 &  0.3 & $-12.9$ &   7.3 &  20.2 &  - & 62.9 $\pm$  0.3 &  424.3 $\pm$    1.0 & - &  - \\ 
    R2 &  0.5 &  $-9.5$ &  16.8 &  26.2 &   - & 71.1 $\pm$  0.5 &  567.5 $\pm$    1.6 & $[+10, +15]$ & 43.3 $\pm$   0.7 \\ 
    \hline
    \multicolumn{9}{c}{SiO $8-7$} \\
    \hline
    B2 &  0.5 & $-16.8$ &   8.6 &  25.4 &  $-2.6$ & 27.6 $\pm$  0.5 &  350.2 $\pm$    1.6 &  $[-8, -15]$ & 87.0 $\pm$   0.7 \\ 
    B3 &  0.4 & $-15.9$ &  11.2 &  27.1 &  $-4.7$ & 25.2 $\pm$  0.4 &  294.2 $\pm$    1.3 &  $[-8, -15]$ & 30.5 $\pm$   0.6 \\ 
    MM1 &  0.3 & $-19.4$ &  15.9 &  35.3 &   3.4 & 11.2 $\pm$  0.3 &  163.2 $\pm$    1.1 & - &  - \\ 
    R2 &  0.4 &  $-7.3$ &  14.2 &  21.5 &   9.0 & 31.8 $\pm$  0.4 &  252.6 $\pm$    1.1 &  $[+10, +15]$ &  32.2 $\pm$   0.5 \\ 
    \hline
    \multicolumn{9}{c}{SO $9_8-8_7$} \\
    \hline
    B2 &  0.1 & $-16.3$ &   6.9 &  23.2 &  $-8.2$ &  4.4 $\pm$  0.1 &   49.4 $\pm$    0.4 &  $[-8, -15]$ &  18.5 $\pm$   0.2 \\ 
    B3 &  0.1 & $-15.1$ &   9.9 &  24.9 &  $-6.4$ & 13.1 $\pm$  0.1 &  140.2 $\pm$    0.5 &  $[-8, -15]$ &  14.6 $\pm$   0.2 \\ 
    MM1 &  0.1 & $-12.9$ &  11.6 &  24.5 &    1.1 & 26.0 $\pm$  0.1 &  238.1 $\pm$    0.4 & - & - \\  
    R2 &  0.1 & $-10.3$ &  12.9 &  23.2 &   7.7 & 11.6 $\pm$  0.1 &  118.8 $\pm$    0.4 & $[+10, +15]$ &   2.2 $\pm$   0.2 \\ 
    \hline
    \multicolumn{9}{c}{SO$_2$ $8_{2,6}-7_{1,7}$} \\
    \hline
    B2 &  0.1 & $-14.1$ &   1.8 &  15.9 & $-10.2$ &  0.4 $\pm$  0.1 &    3.7 $\pm$    0.2 & $[-8, -10]$ &   0.9 $\pm$   0.1 \\ 
    B3 &  0.1 & $-11.5$ &  10.9 &  22.4 &   $-6.5$ & 1.2 $\pm$  0.1 &   11.6 $\pm$    0.3 & $[-8, -10]$ &   0.8 $\pm$   0.1 \\ 
    MM1 &  0.1 &  $-6.0$ &  11.3 &  17.3 &   1.0 &  4.4 $\pm$  0.1 &  - & - &  - \\ 
    R2 &  0.1 &  $-7.2$ &  10.0 &  17.2 &  $-1.2$ &  0.6 $\pm$  0.1 &    5.9 $\pm$    0.2 & $[+8, +10]$ &  0.9 $\pm$   0.1 \\ 
    \hline
    \multicolumn{9}{c}{SO $10_{11}-10_{10}$} \\
    \hline
    MM1 &  0.1 &  $-5.2$ &   4.7 &   9.9 &  - &  0.7 $\pm$  0.1 &   4.2 $\pm$   0.3 & $[-1.9, -3.5]$ &    1.4 $\pm$   0.1 \\ 
            &         &             &          &        &     &                           &                             & $[+1.9, +3.5]$ &    0.8 $\pm$   0.1 \\ 
    \hline     
    \multicolumn{9}{c}{C$^{17}$O $3-2$} \\
    \hline
    MM1 &  0.1 &  $-7.7$ &  10.8 &  18.5 &  $-0.4$ &  7.7 $\pm$  0.1 &   30.7 $\pm$    0.3 & $[-1.9, -3.5]$ &   5.0 $\pm$   0.1 \\ 
            &         &             &          &          &               &                           &                               & $[+1.9, +3.5]$ &  3.2 $\pm$   0.1 \\ 
    \hline     
  \end{tabular}\\
\small 
$^{*}$ velocities are in terms of $V-V_{\rm sys}$ and the error on the velocity values is half of the spectral element, i.e. $\pm 0.2$ \kms\\
$^{a}$ maximum blue- and red-shifted velocity defined as the velocity where the intensity becomes lower than $1 \times$ rms noise\\
$^{b}$ full width at zero intensity $FWZI = | V_{\rm max, r} - V_{\rm max, b} |$\\
$^{c}$ velocity at the emission peak. No value is reported for lines affected by self-absorption, i.e. CO $3-2$ and SO $10_{11}-10_{10}$\\
$^{d}$ main-beam temperature at the emission peak. For lines affected by self-absorption $T_{\rm mb, peak}$ could be a lower limit\\
$^{e}$ line intensity integrated over the whole velocity profile (i.e. over the FWZI)\\
$^{f}$ high velocity range where jet and disk emission are detected\\
$^{g}$ line intensity integrated over HV$_{\rm jet/disk}$. For SO $10_{11}-10_{10}$ and C$^{17}$O $3-2$, $\int T_{\rm mb} dV$ (HV$_{\rm disk})$ is measured for both the blue- and the red-shifted line wings\\
\end{table*}

Figure \ref{fig:SO_vel} shows that as we move to intermediate and high velocities SO $9_8-8_7$ and SO$_2$ $8_{2,6}-7_{1,7}$ blue- and red-shifted emission only slightly overlaps and become more and more collimated. The FWHM is $\sim 0\farcs5$ which translate in an intrinsic width of $\sim 0\farcs2$, i.e. $\sim 90$ AU, after correction for the beam HPBW across the jet ($\sim 0\farcs47$). Assuming an inclination angle $i \sim 4\degr$ \citep{claussen98} the de-projected gas velocity is $V \sim 70-210$ \kms. 
The inferred velocity and width are in perfect agreement with those estimated from SiO $5-4$ emission \citep[see, ][]{codella07,cabrit07b} and indicate that for radial velocities larger than 5 \kms\, SO and SO$_2$ trace the molecular jet.\\

To further investigate the effectiveness of SO and SO$_2$ as tracers of the molecular jet, position-velocity diagrams (PVs) of SO and SO$_2$ emission along and perpendicular to the jet direction (PA$_{\rm jet} = 22\degr$, PA$_{\rm disk}=112\degr$) are extracted and compared with the PVs of CO $3-2$ and SiO $8-7$. 
Figure \ref{fig:pv} shows that the considered tracers have different spatio-kinematical properties.
While SiO traces only jet emission with no contamination from ambient gas, SO and SO$_2$ are dominated by cloud and outflow emission at systemic and low velocities.
Moreover, the LV and IV SO emission shows accelerating ``arms'', from systemic velocity at the source position to $\sim +/- 8$ \kms\, at $\sim \pm 1\arcsec - 1\farcs5$ distance from source, which are not seen in SiO. 
These arms suggest that some SO/SO$_2$-rich gas from the circumstellar envelope or disk may be continuously accelerated to high velocities away from the source, either by interaction with the fast jet or by magneto-centrifugal forces operating on envelope/disk scales \citep[e.g., ][]{ciardi10,panoglou12}.
Also CO 3--2 traces multiple components depending on velocity (the envelope, the outflow, the jet). Moreover, the CO PVs show a strong absorption feature at sligthly red-shifted velocity ($\sim + 1$ \kms). 
Despite the different behaviours of the considered tracers, Figure \ref{fig:so_vs_co_sio} shows that for velocities larger than $\sim +/- 8$ \kms\, the PV diagram of SO along the jet is very similar to that of CO 3--2 and SiO 8--7 suggesting  that at high velocities all tracers probe the same jet component. 
This in turn allows us to compare the emission from SO and CO at high velocities and to estimate the abundances of SO and SO$_2$ in the jet by comparing their column densities with that of CO 3--2 (see Sect. \ref{sect:abu_jet}).

Figure \ref{fig:SO_vel} shows that the SO $9_8-8_7$ and SO$_2$ $8_{2,6}-7_{1,7}$ HV emission has three emission peaks, or knots, marked by black triangles on the SO and SO$_2$ channel maps, two along the blue lobe (B3 and B2) and one along the red lobe (R2). The knots are located at $0\farcs68$, $1\farcs55$, and $1\farcs43$ respectively, i.e. at a de-projected distance from the MM1 protostar of around 300, 700, and 650 AU.
The B2 and R2 SO/SO$_2$ knots are roughly coincident with the B2, R2 knots observed in SiO $5-4$ emission by \citet{codella07} and the SN, SS knots in SiO $8-7$ and CO $3-2$ emission observed by \citet{lee07a}.
1d spectra of SO $9_8-8_7$ and SO$_2$ $8_{2,6}-7_{1,7}$ are extracted at the position of the knots using the same synthetized beam and the obtained intensity profiles ($T_{\rm mb}$ in K versus $(V-V_{\rm sys})$ in \kms) are shown in Figure \ref{fig:spec_jet}. From the spectra we estimate the emission line properties in the knots, i.e.: rms noise in K, maximum blue- and red-shifted velocity ($V_{\rm max, b}$ and $V_{\rm max, r}$ in \kms), full width at zero intensity ($FWZI$ in \kms), velocity at the emission peak ($V_{\rm peak}$ in \kms), main-beam temperature peak ($T_{\rm mb, peak}$ in K), and integrated intensity ($\int T_{\rm mb} dV$ in K \kms)  (see Table \ref{tab:fluxes}).

From the  velocity at the SO emission peak reported in Table \ref{tab:fluxes} we estimate the knot de-projected velocity, $V = V_{\rm peak} / sin(i)$, and derive the knots dynamical timescale, $\tau_{\rm dyn} \sim 15$ years for B3 and $\sim 30$ years for the symmetrically located blue- and red-shifted knots B2 and R2.
The position, distance, velocity and dynamical timescale of the detected SO knots are summarized in Table \ref{tab:knots}.
The inferred dynamical timescales are very low suggesting that the observed SO emission probes very recent ejection events.
Note, however, that these estimates are affected by a large uncertainty due to the assumed inclination angle, the choice of the tracer ($V_{\rm peak}$ varies up to a factor 3 depending on the tracer, either SO, SO$_2$, or SiO), and the broad line profiles (up to $\sim 27$ \kms).
The line broadening may be caused by the jet geometry and/or by internal jet shocks. 
In the first case, if the jet is conical and propagates with constant velocity $V$ at an angle $i$ to the plane of the sky,  the observed maximum blue- and red-shifted velocities (see Table \ref{tab:fluxes}) imply a jet half-opening angle $\theta_{\rm max} \sim 10\degr$\footnote{the jet half-opening angle is $tan \,\theta_{\rm max} = (R - 1)/(R + 1) \times tan(i)$, where R is the algebrical ratio of maximum to minimum radial velocities and $i$ the inclination with respect to the plane of the sky \citep[see ][]{codella07}.}. This in turn implies a knot deprojected velocity $V = V_{\rm max}/sin(i-\theta_{\rm max})$ which is about a factor 2 smaller than the deprojected $V_{\rm peak}$, implying a factor 2 larger timescales  than those reported in Table \ref{tab:knots}.
Note, however, that the observed line profiles are strongly asymmetric  and show broad line wings, unlike the sinthetic profile computed by \citet{kwan88} for a conical jet of constant speed seen close to edge-on (see their Fig. 1, dashed curve for $i = 82.8\degr$ and $2 \theta_{\rm max} = 20\degr$). 
Alternatively, the observed line broadening may be due to an ``internal jet shock'' where fast jet material catches up and shocks slower previous ejecta. In this scenario, over-pressured shocked material is squeezed sideways out of the jet beam, producing an expanding bowshock structure in its wake, which cause the observed line broadening.
The bowshock scenario is favored by \citet{codella07} and further supported by the SiO maps and transverse PVs of \citet{lee08}. In that case, $V_{\rm peak}$ traces the bow wings while the knot velocity along the jet axis is $V = V_{\rm max} / sin(i)$, i.e. around a factor 2 larger than the deprojected $V_{\rm peak}$. This would imply a factor two smaller timescales than those reported in Table \ref{tab:knots}.
We conclude that the estimated knot timescales are affected by an uncertainty of a factor 2-3 depending on the origin of the line broadening and the selected knot tracer.


In Figure \ref{fig:spec_jet} the obtained SO $9_8-8_7$ and SO$_2$ $8_{2,6}-7_{1,7}$ intensity profiles are also compared with the profiles of a typical jet tracer, i.e. SiO $8-7$ \citep{codella07,lee07a}, and a jet/outflow (at high/low velocities respectively) tracer, i.e. CO $3-2$, extracted at the same positions and with very similar synthetized beam (differences between the beams are $\le 0\farcs1$).
The comparison shows that even though the SiO and SO knots are roughly co-spatial SO and SO$_2$ peak at higher velocities than SiO in the blue lobe and at lower velocities in the red one. This stratification in velocity is a typical signature of shock-excited emission as both the abundances of the observed molecular species and the gas physical conditions (density and temperature) undergo strong gradients in the post-shock cooling region, giving rise to chemical segregation.
The different profiles and molecular stratification shown by the symmetrically located knots B2 and R2 suggest different shock conditions in the blue and red lobes within 2\arcsec\, from source. This asymmetry is in contrast with the symmetry observed on larger scales in H$_2$ \citep{zinnecker98}, and suggests that asymmetries may wash out on large timescales.
A detailed comparison of the observed intensity profiles with shock models predictions is out of the scope of this paper and will be presented in Gusdorf et al. (in preparation).

Finally, it should be noted that, even though SO, SO$_2$, SiO, and CO show different behaviours at low velocity (i.e. different spatio-kinematical distribution in the PVs and different line profiles and peak velocities), the maximum blue and red velocities are remarkably similar. This indicates that, as also suggested by the comparison of the line PVs in Figure \ref{fig:so_vs_co_sio}, all the considered lines probe the same jet component at high velocities.


\begin{figure*}
  \begin{centering}
  \includegraphics[width=12.cm]{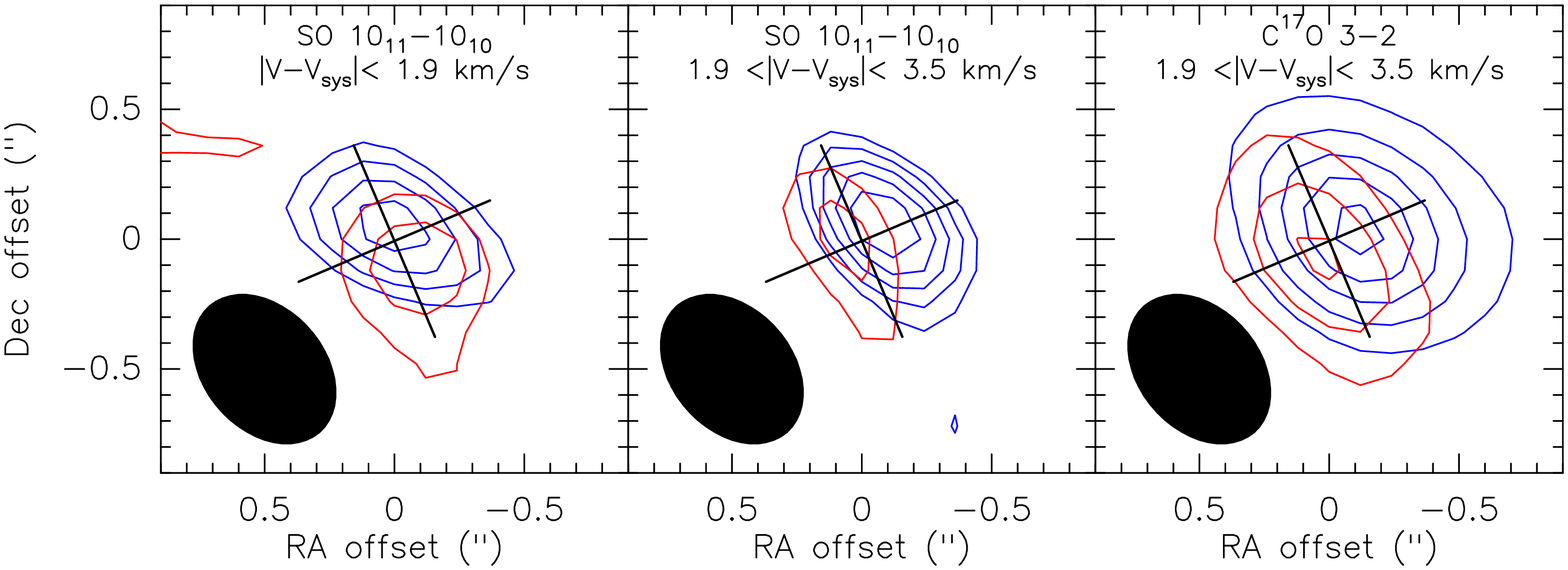}
  \caption{Blue- and red- shifted emission of SO $10_{11}-10_{10}$ at low velocities ($|V-V_{\rm sys} | < 1.9$ \kms, {\it left panel}) and in the higher velocity range where the emission is consistent with a rotating disk (HV$_{\rm disk}$ : $1.9 < |V-V_{\rm sys} | < 3.5$ \kms, {\it middle panel}) is compared with C$^{17}$O $3-2$ emission in  the HV$_{\rm disk}$ range ({\it right panel}, \citealt{codella14b}). The tilted black cross indicates the jet (PA = 22$\degr$) and the disk (PA = 112$\degr$) direction. The ellipse in the bottom-left corner shows the beam HPBW of the line emission maps ($0\farcs65 \times 0\farcs47$, PA = 49$\degr$). First contour at 5$\sigma$ with steps of 5$\sigma$ for C$^{17}$O $3-2$ and at 3$\sigma$ with steps of 1$\sigma$ for SO $10_{11}-10_{10}$.}
  \label{fig:disk}
  \end{centering}
\end{figure*}

\subsection{The flattened rotating envelope}
\label{sect:envelope}

As suggested by \citet{lee07a} while the high-velocity SO 9$_8$-8$_7$ emission is clearly associated with the molecular jet, the low velocity one could arise in the inner rotating envelope and/or in a pesudo-disk.
The bottom panels of Fig. \ref{fig:pv} shows the PVs of the observed emission lines obtained cutting the datacube perpendicular to the jet PA, i.e. along the disk direction (PA$_{\rm disk} = 112\degr$). 
With the exception of SiO, which shows poor emission along the disk, the other lines show a velocity gradient perpendicular to the jet PA of a few \kms\, over a scale $\le 1\arcsec$ (i.e. $\le$450 AU), which suggests rotation in a flattened envelope and/or a disk with the same rotation sense as observed in HCO$^{+}$ by \citet{lee14} and in C$^{17}$O $3-2$ by \citet{codella14b}.

The SO $9_8-8_7$ and SO$_2$ $8_{2,6}-7_{1,7}$ intensity profiles at the position of the MM1 protostar (Fig. \ref{fig:spec_jet}), peak at sligthly redshifted velocity, i.e. $\sim +1$ \kms, where CO $3-2$ shows a deep absorption feature. This could be a signature of infalling gas.
We also note that the SO$_2$ $8_{2,6}-7_{1,7}$ line profile shows two spectral ``bumps'' at blue- and red-shifted velocities, which are not detected in the other tracers.
The PV diagram in Fig. \ref{fig:pv} shows that these spectral features are seen only in a spatially unresolved region towards the MM1 position, and not along the outflow. This suggests that SO$_2$ is blended with two unidentified lines emitted in the inner few 100 AU of the protostellar envelope. These lines are likely produced by Complex Organic Molecules (COMs) evaporated from the icy mantles of dust grains in the warm circumstellar region (T$_{\rm dust} > 100$ K), as observed towards many other low-luminosity protostars \citep[e.g., ][]{maury14}.
The identification and analysis of these lines is out of the scope of this paper.

\subsection{SO 10$_{11}$-10$_{10}$: the compact rotating disk ?}
\label{sect:disk}

\begin{figure}
  \begin{centering}
  \includegraphics[width=\columnwidth]{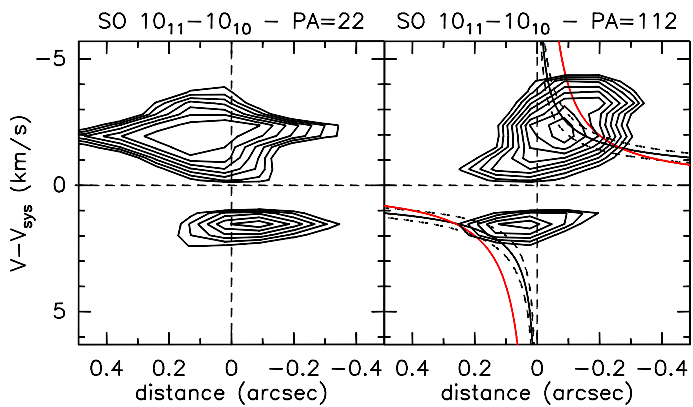}
  \caption{{\it From left to right:} Position-velocity diagram of SO $10_{11}-10_{10}$ obtained along the jet (PA$_{\rm jet} = 22\degr$, {\it left}) and the disk (PA$_{\rm disk} = 112\degr$, {\it right}). Horizonthal and vertical dashed lines mark V$_{\rm sys} = +1.3$ \kms\, and the continuum peak MM1. 
The first contour is at 6$\sigma$ ($1 \sigma = 2$ mJy) with steps of 0.5$\sigma$. 
On the PV along the disk the keplerian curves for $M_{*} = 0.3 \pm 0.1$ \msol\, (solid and dashed black curves) and the r$^{-1}$ curve for infalling gas with angular momentum conservation (red curve) are overplotted.}
  \label{fig:pv_so1010}
  \end{centering}
\end{figure}

Channel maps and position-velocity diagrams along and perpendicular to the jet direction are also obtained for the fainter SO $10_{11}-10_{10}$ line as shown in Figures \ref{fig:disk} and \ref{fig:pv_so1010}. 
At difference with SO  9$_8$-8$_7$ and SO$_2$ $8_{2,6}-7_{1,7}$, the emission in the SO 10$_{11}$-10$_{10}$ line is compact ($\sim 90$ AU scales) and narrow in velocity extending only up to $\pm 5$ \kms\, with respect to systemic.
As shown in Fig. \ref{fig:disk} the red- and blue-shifted emission peaks 
are displaced roughly along the jet direction for velocities $< 1.9$ \kms\, (see left panel). 
However, in the velocity range where C$^{17}$O $3-2$ probes a compact ($\sim 90$ AU) disk rotating around a $0.3\pm0.1$ \msol\, star ($1.9 < |V-V_{\rm sys} | < 3.5$ \kms, \citealt{codella14b}) the SO 10$_{11}$-10$_{10}$ emission peaks are consistently displaced along the disk axis (see middle and right panels).
This suggests that the SO emission at those velocities may also originate in the compact disk.  

The PV diagram perpendicular to the jet axis in Fig. \ref{fig:pv_so1010} supports this scenario as it shows a blue- and a red- lobe which are consistent with emission from a rotating structure.
However,  when cutting the datacube along the jet PA the blue- and red-shifted lobes are consistent with jet emission. 
This indicates that SO 10$_{11}$-10$_{10}$ traces partly rotating gas in a flattened envelope and/or in a disk, and partly the jet. 

\begin{figure}
  \begin{centering}
  \includegraphics[width=8.cm]{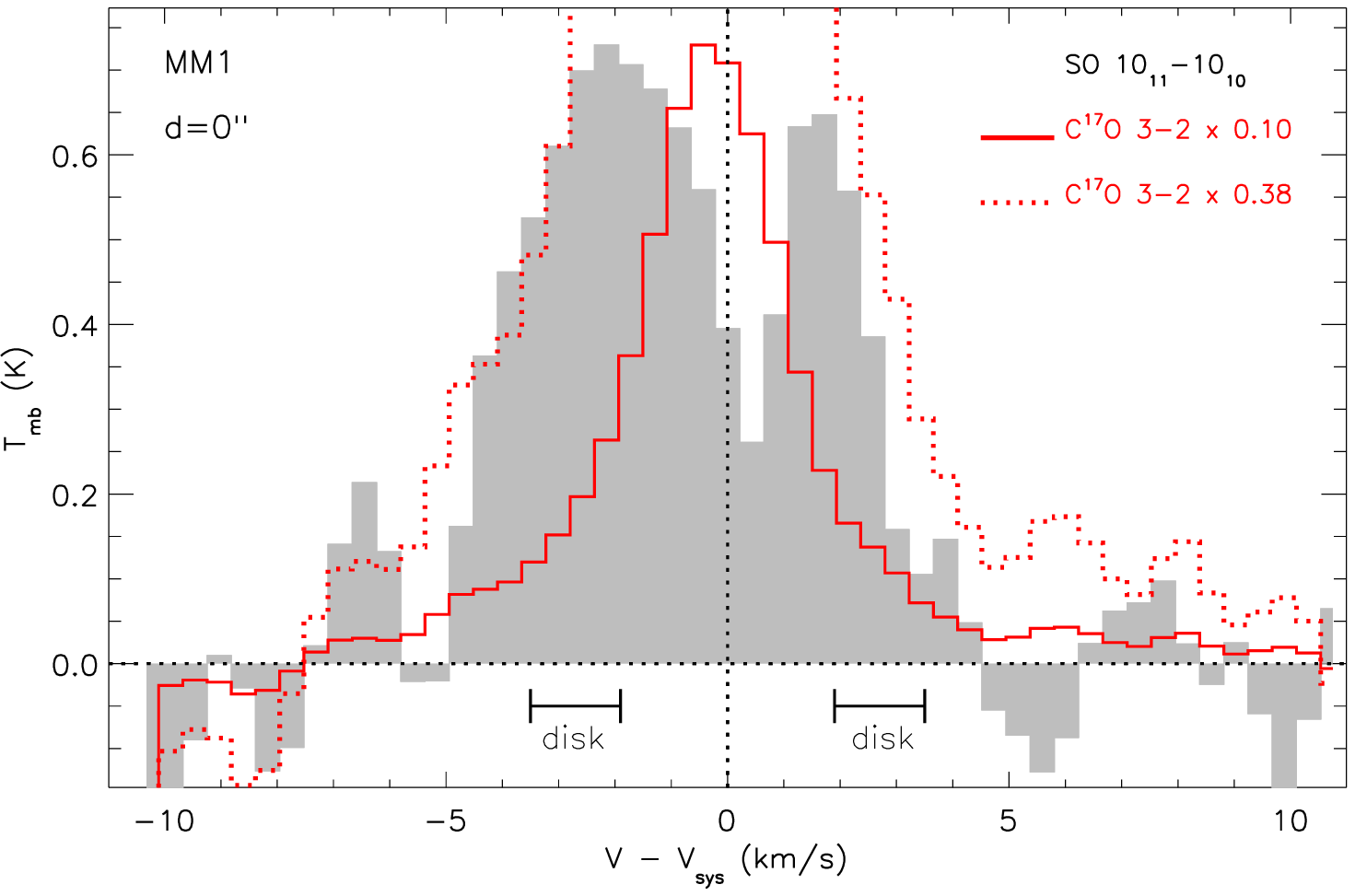}
  \caption{Intensity profiles of SO $10_{11}-10_{10}$ (grey) and C$^{17}$O 3--2 (red) extracted at the source position.
The horizonthal bars in the bottom of the panel indicate the high velocity range (HV$_{\rm disk} = 1.9-3.5$ \kms) where the lines are consistent with a disk origin \citep{codella14b}.
The normalization factor to the SO $10_{11}-10_{10}$ intensity at peak (solid line) and in the blue-shifted HV$_{\rm disk}$ range (dotted line) are labeled on the top of the panel.}
  \label{fig:spec_disk}
  \end{centering}
\end{figure}

In Figure \ref{fig:spec_disk} the  SO 10$_{11}$-10$_{10}$ line intensity profile extracted at the source position is compared with C$^{17}$O $3-2$ analysed by \citet{codella14b}.
In the  $1.9 < |V-V_{\rm sys} | < 3.5$ velocity range the SO 10$_{11}$-10$_{10}$ and C$^{17}$O $3-2$ line profiles are very similar suggesting that at those velocities they are tracing the same disk component.
However, at low velocities the SO 10$_{11}$-10$_{10}$ line intensity profile shows a red-shifted absorption feature at $\sim 0.5$ \kms. 
This may be due to the SO 10$_{11}$-10$_{10}$ higher excitation energy which makes it less affected by the emission from the circumstellar gas and more sensitive to absorption by the infalling gas. As the observed SO 10$_{11}$-10$_{10}$ is compact, the detected absorption feature could probe the infall of the outer disk regions, on much smaller scales than the infalling gas traced by CO $3-2$.

\section{Discussion}
\label{sect:discussion}

\subsection{Searching for jet rotation}
\label{sect:rotation}

Magneto hydro-dynamical (MHD) models predict that the  jet is launched by magneto-centrifugal forces which lift the gas from the disk along the open magnetic field line, accelerating and collimating it.
According to these models the jet rotates around its axis carrying away the excess angular momentum from the disk, thus allowing accretion onto the central star. 
Many recent studies have been devoted to search for systematic velocity asymmetries across the jet axis as a signature of jet rotation, first at optical wavelengths \citep[e.g., ][]{bacciotti02,coffey04} and more recently in the sub-mm and mm range \citep[e.g., ][]{codella07,lee07a,lee07b,lee08,lee09b,launhardt09}.

Tentative rotation measurements have been obtained for HH 212 by \citet{davis00}, \citet{codella07}, and \citet{lee08}.
\citet{davis00} reported a rotation speed of $\sim1.5$ \kms\, at a distance of $\sim230$ AU from the jet axis in the southern H$_2$ knot SK1 located at 2300 AU distance from the source.
However, the velocity asymmetry in the northern lobe would indicate rotation in the opposite sense. 
Moreover, at such large distances from the driving source and from the jet axis the velocity pattern is likely dominated by effects other than rotation, i.e. by interaction with the surrounding cloud in the bow-shock wings, and/or jet wiggling and precession \citep[e.g., ][]{correia09}.
More recent high angular resolution measurements in the inner B2 and R2 knots ($\sim 1\arcsec-2\arcsec$ from source) gives contradictory results: 
\citet{codella07} do not observe transverse velocity shifts above 1 \kms\, in the SiO $5-4$ emission with  $0\farcs78 \times 0\farcs34$ HPBW resolution, 
while higher angular resolution SMA observations of SiO $8-7$ (HPBW $= 0\farcs36 \times 0\farcs33$) indicate a velocity shift at the tip of the two bow-shocks \citep{lee08}.

\begin{figure}
  \begin{centering}
  \includegraphics[width=8.cm]{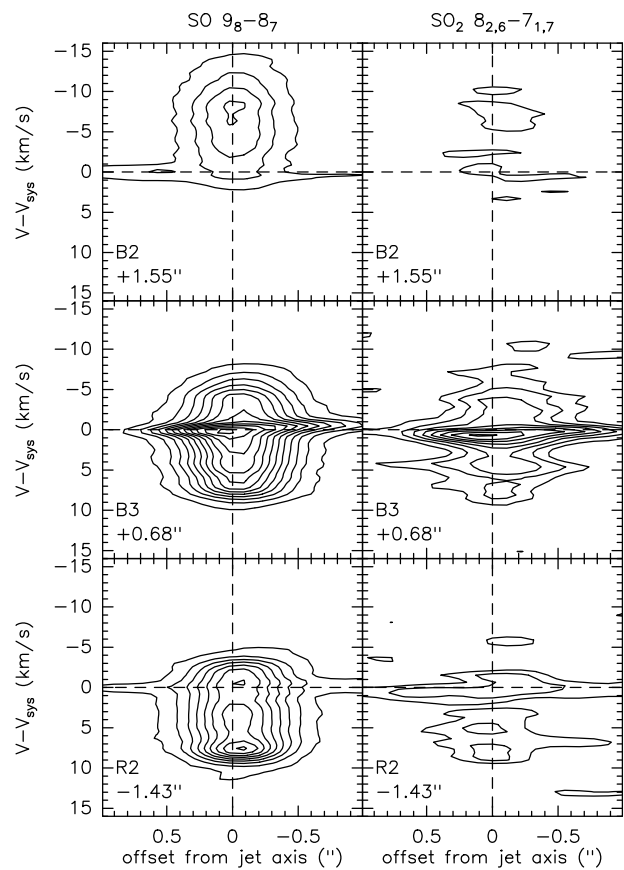}
  \caption{{\it From left to right:} Position-velocity diagrams of SO $9_8-8_7$, and SO$_2$ $8_{2,6}-7_{1,7}$ lines obtained perpendicular to the jet axis (PA = 112$\degr$) at the position of knots B2, B3, and R2. The distance of the knots from the driving MM1 protostar is indicated. The first contour is at 5$\sigma$ ($\sigma = 4$ mJy for SO, and 1.5 mJy for SO$_2$) with steps of 10$\sigma$ (SO), and 5$\sigma$ (SO$_2$).}
  \label{fig:pv_perp_jet}
  \end{centering}
\end{figure}

To test the validity of these tentative rotation measurements we search for velocity asymmetries across the collimated jet by extracting PV diagrams of SO 9$_8$-8$_7$ and SO$_2$ $8_{2,6}-7_{1,7}$ emission perpendicular to the jet axis at the positions of knots B2, B3, and R2 (see Fig. \ref{fig:pv_perp_jet}).
PV diagrams of SiO are not shown as these are at lower resolution with respect to those by \citet{lee08}.
The figure shows no evidence of a velocity gradient across the jet axis, consistent with the high-velocity SO/SO$_2$ jet being practically unresolved transversally. 
Higher angular resolution ($\sim 0\farcs1$) which will be available with the full ALMA array is needed to properly sample the velocity pattern across the jet width and to provide a reliable test of rotation in the HH 212 jet.

In the PV across the B3 knot there is a hint of rotation for the gas close to systemic velocity, which could indicate rotation of the outflow cavity as seen in C$^{34}$S \citep[see][]{codella14b}.

\subsection{SO and SO$_2$ abundance in the jet}
\label{sect:abu_jet}

\begin{figure}
  \begin{centering}
  \includegraphics[width=\columnwidth]{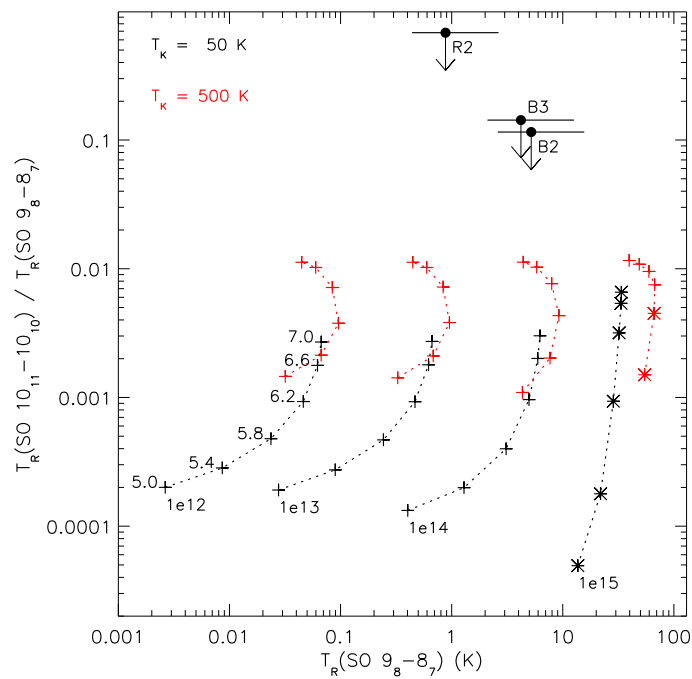}
  \caption{SO line temperature ratio T$_{\rm R}$(SO $10_{11}-10_{10}$)/T$_{\rm R}$(SO $9_8-8_7$) versus intrinsic line temperature T$_{\rm R}$(SO $9_8-8_7$) predicted by RADEX slab models at $T_{\rm K}$ = 50, 500 K (black and red points/curves). Each curve corresponds to the labelled column density $N_{\rm SO}/\Delta V$ increasing from $10^{12}$ to $10^{15}$ cm$^{-2}$ (\kms)$^{-1}$ by factors of 10 (left to right). Each dot along the curves corresponds to the labeled log($n_{H_2}$) (from $10^{5}$ to $10^{7}$ \cmc). Optically thin models are marked by a cross, optically thick one by a star. The black points indicate the upper limit of the SO line ratio and the SO $9_8-8_7$ intensity corrected for beam dilution in the B2, B3 and R2 knots over the HV$_{\rm jet}$ velocity range. 
}
  \label{fig:so_lvg}
  \end{centering}
\end{figure}

\begin{table*}
  \caption[]{\label{tab:abu_jet} Beam-averaged column densities and abundances of SO, SO$_2$, and CO in the jet knots B2, B3, and R2. The column densities are estimated from the SO $9_{8}-8_{7}$,  SO$_2$ $8_{2,6}-7_{1,7}$, and CO $3-2$ line intensities integrated over the indicated high velocity ranges assuming optically thin thermalized emission at $T_{\rm ex} = 50-500$ K. The abundances of SO and SO$_2$ are derived as $X_{\rm specie} = N_{\rm specie} / N_{\rm CO} \times N_{\rm CO} / N_{\rm H_2}$ and assuming $X_{\rm CO} = N_{\rm CO} / N_{\rm H_2} = 10^{-4}$.}
  \begin{tabular}[h]{c|cccc|cccc}
    \hline
    \hline
    Jet    & HV$_{\rm jet}$ & $N_{\rm SO}$               & $N_{\rm CO}$                & $X_{\rm SO}$   & HV$_{\rm jet}$ & $N_{\rm SO_2}$             & $N_{\rm CO}$               & $X_{\rm SO_2}$  \\
    knot & (\kms)          & ($10^{14}$ cm$^{-2}$) & ($10^{17}$ cm$^{-2}$)  & ($10^{-7}$)    &  (\kms)          & ($10^{15}$ cm$^{-2}$) & ($10^{16}$ cm$^{-2}$) &  ($10^{-7}$)     \\ 
    \hline 
    \hline
    B2              & $[-8, -15]$    & $3.0-7.1$ & $\ge 0.4-2.1$$^{*}$ & $\le 7.5-3.3$$^{*}$   & $[-8, -10]$ & $0.1-1.3$ & $\ge 1.8-10.1$$^{*}$ & $\le 4.7-12.4$$^{*}$ \\ 
    B3              & $[-8, -15]$    & $2.3-5.6$ & $0.2-1.2$ & $10.5-4.6$ & $[-8, -10]$ & $0.1-1.1$ & $1.6-9.0$   & $4.7-12.4$ \\
    R2              & $[+10, +15]$ & $0.4-0.8$ & $0.2-1.1$ &  $1.7-0.8$  & $[+8, +10]$ & $0.1-1.3$ & -                 & -                 \\
    \hline     
  \end{tabular}\\
\small
$^{*}$ As CO 3--2 is optically thick in knot B2, the estimated CO column density, $N_{\rm CO}$, is a lower limit, and the SO and SO$_2$ abundances, $X_{\rm SO}$ and $X_{\rm SO_2}$, are upper limits.
\end{table*}

In order to constrain the physical conditions of the gas and the SO abundance in the inner jet knots (B3, B2, and R2) the observed SO line intensities (corrected for beam dilution) and ratios (independent of beam dilution) are compared with the predictions of the statistical equilibrium, one-dimensional radiative transfer code RADEX adopting plane parallel slab geometry \citep{vandertak07}.
Figure \ref{fig:so_lvg} shows the SO line temperature ratio, T$_{\rm R}$(SO $10_{11}-10_{10}$)/T$_{\rm R}$(SO $9_{8}-8_{7}$), versus the SO $9_{8}-8_{7}$ intrinsic line temperature, T$_{\rm R}$(SO $9_{8}-8_{7}$), predicted by RADEX for a reasonable range of kinetic temperatures ($T_{\rm K} = 50 -500$ K) and H$_2$ densities ($n_{H_2}$ from $10^{5}$ to $10^{7}$ \cmc) \citep[e.g., ][]{cabrit07b} and for SO column densities ($N_{\rm SO} / \Delta V$) increasing from 10$^{12}$ to 10$^{15}$ cm$^{-2}$ (\kms)$^{-1}$.
The model predictions are compared with the observed line ratios and line intensities corrected for beam dilution.
As discussed in Sect. \ref{sect:jet} for velocities higher than $\sim 8$ \kms\,  with respect to systemic the SO emission is co-spatial with that of CO 3–2 and SiO 8–7 indicating that SO originate in the collimated molecular jet with no contamination from the surrounding outflow and cloud.
Therefore, the SO ratios and intensity are measured on the jet high-velocity range defined as: HV$_{\rm jet} = [-8,-15]$ \kms\, for knots B3 and B2 and HV$_{\rm jet} = [+10,+15]$ \kms\, for knot R2.
The different velocity range defined for R2 is to avoid the strong absorption feature around $+9$ \kms\, when comparing with the CO 3--2 emission. 
As discussed by \citet{lee07a} and \citet{cabrit12} this feature is present at all positions and may be caused by an extended foreground component fully resolved-out by interferometers. 
The SO $10_{11}-10_{10}$ line is not detected in the jet knots, hence we report the upper limit of the considered SO line ratio. The SO line ratios do not depend on the assumed beam filling factor, as long as both lines trace the same volume of gas.
The observed SO $9_{8}-8_{7}$ line intensity, $T_{\rm mb}$, is corrected for beam dilution, as the jet of $\sim 0\farcs2$ width is broadened by a factor $\sim2$ by beam convolution across the jet.
The errorbars on the SO line intensity in Fig. \ref{fig:so_lvg} accounts for the range of intensity values over HV$_{\rm jet}$, and for possible beam dilution along the jet axis (a factor $\sim 3$ if the knot is circular, see also \citealt{cabrit07b}).

Figure \ref{fig:so_lvg} shows that the avalaible SO observations do not allow constraining the density and temperature of the emitting gas.
However, for the considered temperatures and H$_2$ densities the observations are in agreement with the model predictions for column densities $N_{\rm SO}/\Delta V \sim 10^{13} - 10^{14}$ cm$^{-2}$ (\kms)$^{-1}$. 
For these values of the column density the emission in the considered SO lines is optically thin.

If the density is sufficiently high, i.e. for  $n_{H_2}$ larger than a few 10$^{6}$ \cmc, the emission is also in local termodynamic equilibrium (LTE).
In this case, an estimate of the beam-averaged SO column density in the inner jet knots can be derived from the SO $9_{8}-8_{7}$ line intensity integrated over the high velocities range HV$_{\rm jet}$ assuming optically thin and thermalized emission at $T_{\rm K} \sim T_{\rm ex} = 50 -500$ K.
Following \citet{cabrit12} also the beam-averaged CO column density is derived from the CO 3--2 intensity integrated on the same velocity range as for SO assuming optically thin, LTE emission. The latter assumption is not valid for knot B2 where the ratio between the CO 3--2 and SiO 8--7 emission is $\sim 1$ over the HV$_{\rm jet}$ velocity range. This indicates that both lines are optically thick. Therefore, for knot B2 the derived CO column density is a lower limit.
Then, the abundance of SO is derived as $X_{\rm SO} = X_{\rm CO} \times N_{\rm SO} / N_{\rm CO}$, where the abundance of CO with respect to H$_2$ is assumed to be $X_{\rm CO} = N_{\rm CO} / N_{\rm H_2} = 10^{-4}$.
The main uncertainty on the estimated column densities and abundances is due to the assumption on the gas temperature, therefore we report a range of values corresponding to the temperature values $T_{\rm K} \sim T_{\rm ex} = 50 -500$ K.
Moreover, 
if the density in the jet is lower than a few $10^{6}$ \cmc, the SO $9_{8}-8_{7}$ line is sub-thermally excited implying higher SO column densities. 
In particular, assuming $n_{H_2} = 3 \times 10^{5}$ \cmc, the SO column densities and abundances estimated using RADEX are from a factor 2 (for $T_{\rm K} = 500$ K) to a factor 4 (for $T_{\rm K} = 50$ K) larger than in the LTE-optically thin case.

The SO$_2$ $8_{2,6}-7_{1,7}$ emission is much fainter and covers lower velocities than SO, therefore it is not possible to compare its PV with that of CO 3--2 (see Fig. \ref{fig:pv}).
However, based on the intensity profiles shown in Fig. \ref{fig:spec_jet} we tentatively assume that SO$_2$ and CO 3--2 emission probe the same jet component over the velocity range HV$_{\rm jet} = [-8,-10]$ \kms\, in knots B3 and B2 and that both are optically thin and thermalized.   
Based on these assumptions the column density and abundance of SO$_2$ are inferred by applying the same method as for SO.
The main source of error on the estimated SO$_2$ column densities and abundances is due to the assumed gas temperature, therefore we report a range of values for $T_{\rm K} \sim T_{\rm ex} = 50 -500$ K.
As for SO, if the density in the jet is lower than a few $10^{6}$ \cmc, the SO$_2$ line is sub-thermally excited.
In that case, using RADEX with $n_{\rm H_2} = 3 \times 10^5$ \cmc\, and T$_{\rm ex} = 50-500$ K we obtain a factor 2 to 4 higher column densities and abundances.

The SO, SO$_2$, and CO  line intensities integrated over the high velocities range HV$_{\rm jet}$ in the inner jet knots (B3, B2, and R2) are reported in the last column of Tab. \ref{tab:fluxes}, while the column densities and abundances estimated assuming LTE-optically thin emission are summarized in Tab. \ref{tab:abu_jet}.
The SO abundance in the high velocity jet of HH 212 ($X_{\rm SO} \sim 10^{-7}-2 \times 10^{-6}$) is comparable to that estimated in other Class 0 jets, namely HH 211 \citep{lee10}, the extremely high velocity (EHV) wings of the L1448-mm and IRAS 04166+2706 outflows \citep{tafalla10}, and the B1 and B2 bowshocks at the tip of the L1157 outflow cavity \citep{bachiller97}. On the other hand, the ratio SO$_2$/SO appears variable from jet to jet: we find SO$_2$/SO$\sim 0.5-4$ in HH 212, while it is around 1 in L1157, and $< 0.1-0.3$ in the L1448-mm and IRAS 04166+270 EHV gas.

Assuming a sulfur elemental abundance (S/H)$_{\odot} \sim 1.38 \times 10^{-5}$ \citep{asplund05}, these values indicate that from 1\% up to 40\% of the elemental S  is in the form of SO and SO$_2$ in the HH 212 jet.
This represents an enhancement by a factor 10–100 with respect to SO and SO$_2$ abundances in prestellar cores and protostellar envelopes \citep{bachiller97,tafalla10}.

SO and SO$_2$ enhancements up to $\simeq 10^{-7}-10^{-6}$ are predicted by models of magnetized molecular C-shocks, as a result of neutral-neutral reactions in the shock-heated gas \citep{pineaudesforets93} and sputtering of H$_2$S-rich icy grain mantles \citep{flower94}. The formation of SO$_2$ by oxydation of SO is predicted to occur late in the post-shock, hence the ratio of SO$_2$/SO would be an indicator of shock age (see e.g. Fig 7 of \citealt{flower94}). In the case of HH 212, the ratio SO$_2$/SO~$\ge 0.5$ that we observe seems inconsistent with the very short dynamical timescale of $15-30$ years of the knots. Such a high ratio at early times would require that SO$_2$ is mainly released from grain mantles rather than formed in the gas phase out of SO as assumed in published shock models.
Observations of other sulfur-bearing species, such as CS, OCS, H$_2$S, H$_2$CS, are crucial to constrain the sulfur chemistry and obtain an accurate estimate of the sulphur released in gas-phase in shocks \citep[e.g., ][]{podio14b}.
Alternatively to shocks, high SO and SO$_2$ abundances up to $10^{-6}$ are predicted at the turbulent interface between the outflow and the surrounding cloud material \citep{viti02}. SO and SO$_2$ is also strongly enhanced in an MHD disk wind due to the thermal sublimation of icy mantles near the source and the gradual heating by ambipolar diffusion during MHD acceleration \citep{panoglou12}. Detailed comparison with such models will be the subject of a future paper (Tabone et al., in prep).

\subsection{Jet physical and dynamical properties: H$_2$ density and mass loss rate}
\label{sect:mjet}

\begin{table*}
  \caption[]{\label{tab:knots} Properties of the knots along the HH 212 jet: position offset (RA and dec in \arcsec), on-sky and deprojected distance from the MM1 protostar ($d_{\rm tan}$ in \arcsec\, and $d$ in AU), deprojected peak velocity ($V$ in \kms), dynamical timescale ($\tau_{\rm dyn}$ in yr), high velocity range on which the jet properties are estimated (HV$_{\rm jet}$ in \kms),  beam-averaged H$_2$ column density ($N_{\rm H_2}$ in cm$^{-2}$), H$_2$ volume density ($n_{\rm H_2}$ in cm$^{-3}$), and mass loss rate ($\dot{M}_{\rm jet}$ in \msolyr).}
  \begin{tabular}[h]{ccccccccccc}
    \hline
    \hline
    Knot & RA              & Dec & $d_{\rm tan}$  & $d$ & $V$ & $\tau_{\rm dyn}$ & HV$_{\rm jet}$ & $N_{\rm H_2}$ & $n_{\rm H_2}$ & $\dot{M}_{\rm jet}$ \\
            & ($\arcsec$) & ($\arcsec$) & $('')$ & (AU) & (\kms) & (yr)               & (\kms)           & (cm$^{-2}$)   & (cm$^{-3}$)   & (\msolyr) \\
    \hline 
    \hline
    B2 & $+ 0.63$ & $+ 1.42$ & $+1.55$ & 700 & $-118$ & 30 &    $[-8, -15]$ & $\ge 0.4-2.1 \times 10^{21}$$^{*}$ & $\ge 0.6-3.1 \times 10^{6}$$^{*}$ & $\ge 0.4-1.9 \times 10^{-6}$$^{*}$ \\
    B3 & $+ 0.25$ & $+ 0.63$ & $+0.68$ & 300 &   $-92$ & 15 &    $[-8, -15]$ & $0.2-1.2 \times 10^{21}$ & $0.3-1.8 \times 10^{6}$ & $0.2-1.1 \times 10^{-6}$ \\
    R2 & $-0.59$  & $-1.30$   & $-1.43$ & 650 & $+110$ & 30 & $[+10, +15]$ &$0.2-1.1 \times 10^{21}$ & $0.3-1.6 \times 10^{6}$ & $0.2-1.0 \times 10^{-6}$ \\
    \hline     
  \end{tabular} \\
\small
$^{*}$ As CO 3--2 is optically thick in knot B2, the estimated CO column density, $N_{\rm CO}$, and hence H$_2$ column and volume density, $N_{\rm H_2}$ and $n_{\rm H_2}$, and mass loss rate, $\dot{M}_{\rm jet}$, are lower limits.

\end{table*}

From the beam-averaged CO column density estimated in the HV$_{\rm jet}$ range, $N_{\rm CO}$ (see Tab. \ref{tab:abu_jet}), one can infer the H$_2$ column and volume density in the knots along the jet, $N_{\rm H_2}$ and $n_{\rm H_2}$, and the jet mass loss rate, $\dot{M}_{\rm jet}$.
The determination of these quantities relies on the assumption that CO 3--2 emission at high velocity is optically thin and traces all the emitting mass in the beam. As CO 3--2 in knot B2 is optically thick the $N_{\rm H_2}$, $n_{\rm H_2}$, and $\dot{M}_{\rm jet}$ values inferred for this knot are lower limits.
First, the beam-averaged H$_2$ column density is derived assuming $X_{\rm CO} = N_{\rm CO} / N_{\rm H_2} = 10^{-4}$.
Then, the H$_2$ density is derived assuming that the jet is a cylinder uniformely filled by the observed gas. Hence,  $n_{\rm H_2} = (HPBW_{\rm tr} / 2 R_{\rm jet}) \times N_{\rm H_2} / 2 R_{\rm jet}$, where $2 R_{\rm jet}$ is the jet width ($\sim 90$ AU) and the factor $(HPBW_{\rm tr} / 2 R_{\rm jet})$ accounts for the beam dilution of the $\sim 90$ AU jet emission in the transverse beam direction.
Finally, the mass loss rate is estimated assuming that the mass in the jet flows at constant density and speed along the jet axis over the beam length. The latter  assumption is not fulfilled if the gas in the knots is highly compressed by shocks, therefore following \citet{lee07a,lee07b} we correct the mass loss rate for a compression factor of $\sim 3$.
Based on the above assumptions, $\dot{M}_{\rm jet}$ is calculated as $\dot{M}_{\rm jet} = 1/3 \times m_{\rm H_2} \times (N_{\rm CO} / X_{\rm CO}) \times HPBW_{\rm tr} \times V_{\rm jet}$, where $V_{\rm jet}$ is the de-projected jet velocity in the considered HV$_{\rm jet}$ range ($V_{\rm jet} \sim 165$ \kms) and $HPBW_{\rm tr}$ is the beam size across the jet width ($HPBW_{\rm tr} = 0\farcs47$).
Note that the correction for compression is uncertain, as it depends on the unknown shock parameters (magnetic field and shock speed), therefore the $\dot{M}_{\rm jet}$ values are affected by a factor $\sim 3$ uncertainty.
Moreover, the estimated mass loss rate accounts only for the mass transported by the molecular jet component at high velocity ($|V - V_{\rm sys}| > 8$ \kms).
The total jet mass loss rate may be up to a factor of a few higher than the estimated value as the low velocity component can considerably contribute to the total mass loss rate if it traces a slow wide angle wind rather than entrained ambient material (see, e.g., \citealt{maurri14}).
If the jet is partially ionized the atomic gas may further contribute to the mass loss rate. 
However, Spitzer and {\it Herschel} observations of molecular jets driven by Class 0 sources indicate that the mass flux transported by the atomic component is about 10 times smaller than  that carried by the molecular gas \citep{dionatos09,dionatos10,nisini15}.
Hence,  the atomic component of HH 212, if any, should not significatly contribute to the total jet mass loss rate.

The inferred values of  $N_{\rm H_2}$, $n_{\rm H_2}$, and $\dot{M}_{\rm jet}$ for the B2, B3, and R2 knots are summarized in Tab. \ref{tab:knots}.
These quantities are affected by the uncertainty on the $N_{\rm CO}$ estimates, hence a range of values is given depending on the  temperature assumed to derive $N_{\rm CO}$ ($T_{\rm ex} = 50, 500$ K).
As explained above, the $\dot{M}_{\rm jet}$ values inferred for the molecular high velocity jet component are a lower limit to the total jet mass loss rate.  Therefore, the mass flux transported by the jet  ($\dot{M}_{\rm jet} \ge 0.2 - 2 \times 10^{-6}$ \msolyr) is at least 3\% to 33\% of the envelope infall rate ($\dot{M}_{\rm inf} \sim 6 \times 10^{-6}$ \msolyr, \citealt{lee06}), indicating an high jet efficiency ($\dot{M}_{\rm jet}/\dot{M}_{\rm inf} \ge 0.03-0.3$), in agreement with the young age of the source.
Note that the mass accretion rate from the disk onto the central source may be higher than the envelope infall rate.
However, there are no estimates of the disk mass accretion rate for HH 212 as Class 0 sources like this one are too deeply embedded in their parental envelope to enable the use of well-calibrated accretion tracers such as, e.g., the Br$\gamma$ line.

\subsection{Disk properties: mass and SO abundance}
\label{sect:abu_disk}

\begin{figure}
  \begin{centering}
  \includegraphics[width=\columnwidth]{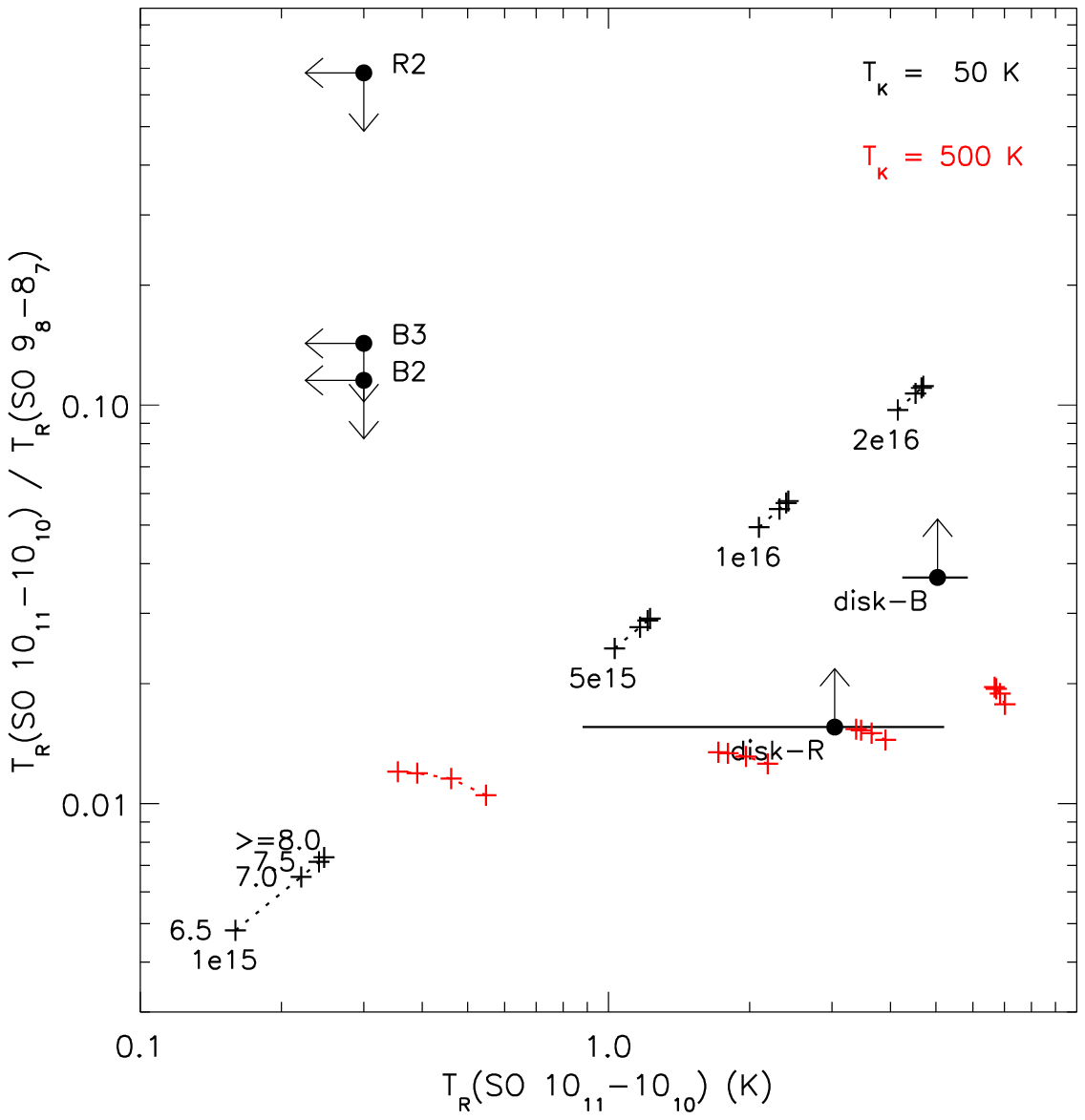}
  \caption{SO line temperature ratio T$_{\rm R}$(SO $10_{11}-10_{10}$)/T$_{\rm R}$(SO $9_8-8_7$) versus intrinsic line temperature T$_{\rm R}$(SO $10_{11}-10_{10}$) predicted by RADEX slab models at $T_{\rm K}$ = 50, 500 K (black and red points/curves). Each curve corresponds to the labelled column density $N_{\rm SO}/\Delta V$ increasing from $10^{15}$ to $2 \times 10^{16}$ cm$^{-2}$ (\kms)$^{-1}$ going from left to right. Each dot along the curves corresponds to the labeled log($n_{H_2}$) (from $10^{6.5}$ to $\ge 10^{8}$ \cmc). Models for which the emission in the SO $10_{11}-10_{10}$ is optically thin are marked by a cross, optically thick ones by a star. The black points indicate the SO line ratio and the SO $10_{11}-10_{10}$ line intensity (or their lower/upper limits) corrected for beam dilution in the B2, B3 and R2 knots and in the disk blue and red lobes. 
}
  \label{fig:so10_lvg}
  \end{centering}
\end{figure}

\begin{table}
  \caption[]{\label{tab:abu_disk} Beam-averaged column densities and abundances of SO and C$^{17}$O in the disk. The column densities are derived by integrating the line intensity of SO $10_{11}-10_{10}$ and C$^{17}$O $3-2$ over the velocity range defined by \citet{codella14b} (HV$_{\rm disk}$: $| V-V_{\rm sys} | = 1.9-3.5$ \kms). We assume optically thin emission at LTE with $T_{\rm ex} = 50-500$ K. The abundance of SO is derived as $X_{\rm specie} = N_{\rm specie} / N_{\rm C^{17}O} \times  N_{\rm C^{17}O} /  N_{\rm CO} \times  N_{\rm CO} /  N_{\rm H_2}$ and assuming $N_{\rm C^{17}O} / N_{\rm CO} = 1/ 1792$ and $X_{\rm CO} = N_{\rm CO} / N_{\rm H_2} = 10^{-4}$ \citep{wilson94}.}
  \begin{tabular}[h]{ccccc}
    \hline
    \hline
    Disk & HV$_{\rm disk}$ & $N_{\rm SO}$                & $N_{\rm C^{17}O}$         & $X_{\rm SO}$ \\
    lobe  & (kms)              & ($10^{15}$ cm$^{-2}$) & ($10^{16}$ cm$^{-2}$) & ($10^{-7}$) \\ 
    \hline 
    \hline
    blue & $[-1.9, -3.5]$ & $6.0-4.6$ & $0.2-1.3$ & $1.4 - 0.2$\\
    red  & $[+1.9, +3.5]$ & $3.4-2.6$ & $0.2-0.9$ & $1.2 - 0.2$\\
   \hline     
  \end{tabular}\\
\small
\end{table}

An analysis similar to that performed in Sect. \ref{sect:abu_jet} is applied to constrain the gas physical conditions and the SO abundance in the disk.
The observed SO $10_{11}-10_{10}$ line intensity and the SO $10_{11}-10_{10}$/SO $9_{8}-8_{7}$ ratio in the blue and red disk lobes (HV$_{\rm disk} : |V-V_{\rm sys}| = 1.9-3.5$ \kms, \citealt{codella14b}) are compared with the predictions of RADEX computed for $T_{\rm K} = 50 -500$ K, H$_2$ densities larger than $3 \times 10^{6}$ \cmc\, and SO column densities $N_{\rm SO} / \Delta V = 10^{15}, 5 \times 10^{15},  10^{16}, 2 \times 10^{16}$ cm$^{-2}$ (\kms)$^{-1}$.
The observed SO $10_{11}-10_{10}$ line intensity, $T_{\rm mb}$, is corrected for beam dilution, as the emission from the compact disk lobes ($R_{\rm disk} \sim 90$ AU, i.e. $0\farcs2$) is strongly diluted in the $ 0\farcs65 \times 0\farcs47$ beam (dilution factor $\sim 8$). The SO $9_{8}-8_{7}$ emission at low velocities is heavily dominated by the outflow emission, hence the observed SO $10_{11}-10_{10}$/SO $9_{8}-8_{7}$ ratio is a lower limit to the SO line ratio in the disk.
Figure \ref{fig:so10_lvg} shows that, to reproduce the bright  SO $10_{11}-10_{10}$ emission, column densities of $N_{\rm SO}/\Delta V \sim 1-2 \times 10^{16}$ cm$^{-2}$ (\kms)$^{-1}$ are required, i.e. two orders of magnitude larger than what estimated in the jet ($N_{\rm SO}/\Delta V \sim 10^{14}$). 
In the considered models the emission in the SO $9_{8}-8_{7}$ line is optically thick (see also Fig. \ref{fig:so_lvg}), while the emission in the SO $10_{11}-10_{10}$ line is still optically thin.
Moreover, volume densities are expected to be larger than $10^{7}$ \cmc\, in the disk \citep{lee14}, therefore we can assume that the line emission is thermalized.

As the emission in the SO 10$_{11}$-10$_{10}$ is optically thin and thermalized, we estimate the SO column density in the disk by integrating the SO 10$_{11}$-10$_{10}$ line intensity on the velocity interval where the emission is consistent with a disk origin, HV$_{\rm disk}$, and assuming $T_{\rm K} \sim T_{\rm ex} = 50 - 500$ K.  
Given its low abundance ($X_{\rm C^{17}O} = X_{\rm CO} / 1792$, \citealt{wilson94}) and critical density the emission in the C$^{17}$O 3--2 line is also assumed to be optically thin and thermalized at $T_{\rm K} \sim T_{\rm ex} = 50 - 500$ K and the C$^{17}$O column density is inferred from  the C$^{17}$O 3--2 line intensity integrated on HV$_{\rm disk}$ as for SO.
The  SO 10$_{11}$-10$_{10}$ and C$^{17}$O 3--2 line intensity integrated on HV$_{\rm disk}$ are reported in the last column of  Table \ref{tab:fluxes}, while the estimated beam-averaged SO and C$^{17}$O column densities are summarized in Table \ref{tab:abu_disk}. 

From the beam-averaged C$^{17}$O column density, $N_{\rm C^{17}O}$, we estimate the H$_2$ column and volume density in the disk and its mass.
The beam-averaged H$_2$ column density is derived assuming $X_{\rm C^{17}O} = X_{\rm CO} / 1792$ and $X_{\rm CO} = N_{\rm CO} / N_{\rm H_2} = 10^{-4}$.
Then, as the disk is seen almost edge-on, the H$_2$ density is $8 \times N_{\rm H_2} / R_{\rm disk}$, where $R_{\rm disk} \sim 90$ AU and the factor $8$ accounts for the beam dilution.
Finally, the disk mass is obtained by integrating the beam-averaged C$^{17}$O column density over the beam area, $A_{\rm beam}$, and summing over the blue and red disk lobes: $M_{\rm disk} \sim \left( 1.4 \times m_{\rm H_2} \times (N_{\rm C^{17}O} / X_{\rm C^{17}O}) \times A_{\rm beam} \right)_{\rm blue + red}$, where the factor $1.4$ accounts for the mass in the form of Helium. 

Depending on the adopted temperature value, the estimated beam-averaged H$_2$ column density and H$_2$ volume density in the disk are $N_{\rm H_2} \sim 0.4 - 2 \times 10^{23}$ cm$^{-2}$ and $n_{\rm H_2} \sim 0.2 - 1 \times 10^{9}$ \cmc, where the lower and higher values correspond to $T_{\rm ex} = 50$, $500$ K, respectively.
The disk mass turns out to be $\sim 0.002 - 0.013$ \msol. 
This value is close to the disk mass  of $0.014$ \msol\, estimated by \citet{lee14} from HCO$^{+}$ $4-3$.
Hence, if C$^{17}$O is tracing the same disk layer as HCO$^{+}$, it is indeed optically thin and all in the gas phase in this region. Note that more mass could be hidden in the dark disk mid plane where molecules are frozen onto dust grains, so our estimate is only a lower limit to the total disk mass, i.e. $M_{\rm disk} \ge 0.002 - 0.013$ \msol.

Finally, we estimate the SO gas-phase abundance in the disk by assuming that SO 10$_{11}$-10$_{10}$ and C$^{17}$O 3--2 originate from the same disk layer in the $1.9-3.5$ \kms\, velocity range and that C$^{17}$O is all in gas-phase in the considered disk-emitting region. 
Based on the above assumptions $X_{\rm SO} = X_{\rm C^{17}O} \times N_{\rm SO} / N_{\rm C^{17}O} \sim 10^{-8} - 10^{-7}$.
This estimate would remain correct even if C$^{17}$O is depleted onto grains, as long as SO and C$^{17}$O are depleted by the same amount (and both optically thin).
The estimated SO column density and abundance are much higher than what observed in evolved protoplanetary disks ($N_{\rm SO} \le 10^{13}$ cm$^{-2}$, $X_{\rm SO} \le 10^{-11}$, \citealt{fuente10,dutrey11}) and may be due to an SO enhancement in the accretion shock at the envelope-disk interface as suggested by \citet{lee14,sakai14}.
Alternatively, an high SO abundance (up to $10^{-6}$) is predicted by self-gravitating disk models due to the shocks occuring in the disk spirals caused by gravitational instabilities \citep{ilee11,douglas13}. 
Since the central star has a mass of only $0.2-0.3$ \msol\, \citep{lee14,codella14b} and our estimates of the disk mass do not include the mass reservoir frozen onto grains, such a scenario is not excluded.
However, the presented ALMA observations do not allow us to distinguish between these scenarios. The angular resolution ($\sim 0\farcs5$) is too low to fully resolve the disk, therefore the SO 10$_{11}$-10$_{10}$ line profile is averaged over a large portion of the disk impeding to detect SO enhancements as a function of velocity. Moreover, an edge-on view is not favorable to search for SO enhancements in a gravitationally unstable disk as each line of sight would cross many spiral arms. Such a scenario could however be tested for less inclined Class 0 disks with higher resolution ALMA observations.

.

\section{Conclusions}
\label{sect:conclusions}

HH 212 is the prototype of the Class 0 jet and it has been the subject of a number of high angular resolution studies at mm wavelengths performed with the SMA and the PdBI \citep{lee06,lee07a,lee08,codella07,cabrit07b,cabrit12}.
The high sensitivity of the ALMA Cycle 0 observations allow us for the first time to image not only the envelope, the outflow, and the jet but also the inner compact disk (see also \citealt{lee14, codella14b}).
In particular the presented analysis of SO and SO$_2$ emission show that these molecules probe
different components in the inner $\sim 1300$ AU of the HH 212 protostellar system : 
(i) the circumstellar gas and the X-shaped cavity walls at systemic velocity; 
(ii) the poorly collimated gas (width $\sim 135$ AU) likely swept out by the high-velocity molecular jet at low velocities;
(iii) the collimated ($\sim 90$ AU) and fast ($\sim 100 - 200$ \kms) molecular jet at intermediate and high velocities;
(iv) a compact structure with the blue- and the red-shifted peaks displaced roughly on the equatorial plane, which is consistent with an origin in the $90$ AU disk rotating around a $\sim 0.2-0.3$ \msol\, star previously detected by \citet{lee14} and \citet{codella14b}, in the high-excitation SO $10_{11}-10_{10}$ line.

The high-velocity molecular jet detected in SO, SO$_2$, CO, and SiO lines is dense ($\sim 10^5 - 10^6$ \cmc) and transport a mass loss rate $\dot{M}_{\rm jet} \ge 0.2-2 \times 10^{-6}$ \msolyr, indicating a high ejection efficiency ($\dot{M}_{\rm jet} / \dot{M}_{\rm inf} \ge 0.03-0.3$, for an infall rate of $\sim 6 \times 10^{-6}$ \msolyr, \citealt{lee06}).
The abundance of SO in the jet is $X_{\rm SO} \sim 10^{-7} - 10^{-6}$, similar to what estimated in other Class 0 outflows and jets \citep[e.g., ][]{bachiller97,tafalla10,lee10}, while $X_{\rm SO_2} / X_{\rm SO} \ge 0.5$. This indicates that between 1\% and 40\% of the elemental S has been converted into SO and SO$_2$ in the shocks occuring along the jet or at the outflow-cloud interface \citep{pineaudesforets93,flower94,viti02} and/or at the base of the wind due to ambipolar diffusion \citep{panoglou12}.

From C$^{17}$O $3-2$ emission we estimate a disk mass $\ge 0.002 - 0.013$ \msol, which  is consistent with the estimate derived from HCO$^{+}$ by \citet{lee14}.
The SO abundance in the disk ($X_{\rm SO} \sim 10^{-8} - 10^{-7}$) is $3-4$ orders of magnitude larger than what estimated in evolved protoplanetary disks \citep{fuente10,dutrey11}.
Such an SO enhancement may be produced in the accretion shock occuring at the interface between the envelope and the disk as suggested by \citet{lee14} and \citet{sakai14}, or in spiral shocks if the disk is partly gravitationally unstable \citep{ilee11,douglas13}.
Higher angular resolution and sensitivity observations are required to address the exact origin of the detected high-excitation SO emission.

\begin{acknowledgements}
LP has received funding from the European Union Seventh Framework Programme (FP7/2007-2013) under grant agreement n. 267251.
This work was partly supported by the Italian Ministero dell’Istruzione, Universit\'a e Ricerca (MIUR) through the grants Progetti Premiali 2012 - iALMA and PRIN 2013 - JEDI.
RB and MT gratefully acknowledge partial support from Spanish MINECO under Grant FIS2012-32096.
\end{acknowledgements}

\bibliographystyle{aa} 

\end{document}